\documentstyle[epsfig,aps,amssymb]{revtex}
\def\Rc{\check{R}}
 1
\begin{document}

\title{Results on the symmetries of integrable fermionic models on chains}
\author{Fabrizio Dolcini,
and Arianna Montorsi \footnote{e-mail: fdolcini@athena.polito.it \, \, ;
\, \, montorsi@athena.polito.it}}
\address{Dipartimento di Fisica and Unit\`a INFM, Politecnico di Torino,
I-10129 Torino, Italy}
\date{\today}
\maketitle
\begin{abstract}
We investigate integrable fermionic models within the scheme of the graded
Quantum Inverse Scattering Method, and prove
that any symmetry imposed on the solution of the Yang-Baxter Equation
reflects on the constants of motion
of the model; generalizations with respect to known results are discussed.
This theorem is shown to be very effective when combined with
the Polynomial $\Rc$-matrix Technique (PRT): we apply
both of them to the study of the extended Hubbard models, for which we
find all the subcases enjoying several kinds of
(super)symmetries. In particular, we derive a geometrical construction
expressing any $gl(2,1)$-invariant model as a linear combination of EKS
and U-supersymmetric models. Furtherly, we use the PRT to obtain 32 integrable
$so(4)$-invariant models. By joint use of the Sutherland's Species technique
and $\eta$-pairs construction we propose a general method to derive their
physical features, and we provide some explicit results.
\end{abstract}

1999 PACS number(s): 71.10.Fd; 71.27.+a; 75.10.Lp
\\Keywords: Quantum Inverse Scattering Method, Yang-Baxter Equation,
Extended Hubbard models, Symmetries
%%%%%%%%%%%%%%%%%%%%%%%%%%%%
\section{Introduction}
\noindent The issue of finding exact solutions for quantum models
is strictly connected with the notion of {\it integrability} of
their Hamiltonian $H$. Integrability actually consists in the
possibility of finding a complete set of commuting observables,
so that the eigenstates of $H$ can be univoquely characterized by
the values of the quantum numbers. Within this subject, the
Quantum Inverse Scattering Method \cite{KUL,FAD,PU-ZH,KBI} is a
very powerful tool providing sets of commuting operators; it is
based on the Yang-Baxter Equation (YBE), a functional equation
for a ${\Bbb C}$-number matrix (the $R$-matrix), which can be
thought of as a factorizability condition for the scattering
matrix. Such techniques were originally applied to study
non-linear PDE as well as spin systems, for which some remarkable
results have been obtained. As to fermionic problems, the
traditional approach was to use the Jordan-Wigner \cite{OWA,SHA}
transformation to map them into spin systems, and then to study
the latter by means of the QISM.
\\More recently, new methods have been worked out to {\it directly}
apply the QISM to systems of fermions: in the literature two main
different approaches can be found to this aim; the first one makes
use of algebraic representation theory to find solutions
of the YBE, introducing a grading that accounts for the fact that the
operators of fermionic systems can either commute or anticommute.
The second method (fermionic $R$-matrix) is quite direct: it consists
in dealing with an {\it operator-valued} YBE (instead of a
${\Bbb C}$-number YBE), so that everything is calculated directly in
terms of fermionic operators. Despite the remarkable results obtained
within each approach, interesting questions are still open, which we
shall investigate in this paper.
\\In the first instance, the passage `${\Bbb C}$-number'
$\leftrightarrow$ `fermionic operators' should be clarified in
order to use in a complementary way the two approaches cited
above: in fact the algebraic methods are very powerful in
providing solutions of the ${\Bbb C}$-number YBE, but their
results are not always `translated' into fermionic operators, so
that the Hamiltonian and the constants of motion are not easily
readable; on the other hand, the fermionic $R$-matrix
formulation, working with operators instead of ${\Bbb
C}$-numbers, is more direct but quite cumbersome on a
computational point of view. In the preliminary sections of this
paper we therefore clarify the mutual relationship between these
two approaches: sec.II and sec.III provide systematic methods to
represent fermionic operators with matrices and to derive fermion
models from the ${\Bbb C}$-number YBE. In sec. IV some particular
remarks are made on the structure of the graded tensor product,
which is crucial to this interplay.
\\Secondly, a quite important open problem is concerned with the
symmetries. Indeed, even if the algebraic methods allow to classify
the solutions of the YBE according to several kinds of
(super)symmetries, no much attention has been paid to investigate
the effect that these (local)
symmetries have on the constants of motion that the QISM determines.
Although some results have been obtained with the fermionic $R$-matrix
approach in a particular
case~\cite{USW}, just a few attempts \cite{GOMU} have been made
to find general results on this issue, or to develop them. We point
out that knowing the symmetries of
the constants of motion of exact results is important when developing
numerical approaches to study perturbations on such exact models;
in fact, by fine-tuning the perturbation, one can control what happens
to the conserved quantities and realize how many of them get broken.
This is crucial because in quantum mechanics it is
not that easy to have an analogue of the classical Poincar\'e sections
for studying the break-up of invariant toruses.
Moreover, the investigation of
symmetries is also very useful to determine the physical features of
a model: indeed the few {\it exact} phase diagrams of
interacting fermion models that we know\cite{EKSbook}, have been derived
just exploiting the symmetry properties. It is therefore quite important,
once a given symmetry has been realized to provide interesting
physical insights, to find all the fermionic models (within a certain
general class of Hamiltonians) that fulfill it, in order to deduce the
properties of further models. Also, it is important to know how
many and what kind of free parameters the imposition of a certain
symmetry allows on a given class of models; in
fact, even if the exact solution is not possible in general for all
the values of the above free parameters, one can select which is the
most suitable parameter for the investigation of a given physical
feature, and see what happens when varing it while keeping the others
fixed, the symmetry being always preserved.
\\We therefore devote all the central
sections of our paper to present results on the symmetries. In particular
in sec.V, concerned with the symmetries
of the YBE equation itself, we extend the invariance properties
of the YBE to the {\it graded} similarity transformations, which are
the actually useful similarity transformations when dealing with
fermionic systems. In section VI we prove a theorem assuring
that the local symmetries imposed on the $\Rc$-matrix
directly reflect on the constants of motion; here we shall generalize
some known results by considering generic symmetries and  possible
additional signs in the generators of neighboring sites. In
sec.VII we investigate the complemantary question, namely
the conditions under which the symmetries of the Hamiltonian reflect
onto the $\Rc$-matrix; the Polymonial $\Rc$-matrix Technique~(PRT)
\cite{DOMO1} is shown to provide interesting general answers to this
issue. Finally, the two  results of sec.VI and VII are
combined together to investigate the specific case of the Extended Hubbard
Models (EHM). More precisely, in sec.VIII we exploit the matrix
representation of sec.II to find all the EHM that fulfill
several kinds of (super)symmetries (in particular, we provide a
geometrical construction allowing to express any
$gl(2,1)$-invariant model as a linear combination of the EKS
model\cite{EKS4,EKS} and the U-supersymmetric model\cite{BEFR,RAMA2});
furtherly, in analyzing the integrable subcases
of each of the above symmetry classes, we realize that for most of them
the $\Rc$-matrix is a first or second order polynomial; hence, we
are able to deduce the symmetry properties of the constants of motion for
such models ({\it e.g.} for EKS and U-supersymmetric models).
\\Finally, in sec. IX we focus on the $so(4)$ symmetries
which are proved to be particularly interesting in condensed matter
physics. By means of the PRT we find all the models that are derivable
by first degree polynomial $\Rc$-matrices, and through the theorem
of section~VI we show that the constants of motion are
$so(4)$-invariant as well. All the models we find turn out to act as
Generalized Permutators; this allows us to exploit the Sutherland's Species
technique, and to  propose a general scheme for finding the spectrum
and the ground state phase diagram. The joint use of the $\eta$-pairs
construction (based on the $so(4)$-symmetry), leads to determine the
eigenstates and the correlation functions for these models, even when
arbitrary Coulomb repulsion and filling are allowed.
\label{sec-1}
%%%%%%%%%%%%%%%%%%%%%%%%%%%%%%%%%%%%%%%%%%%%%%%%%%%%%%%%
%%%%%%%%%%%%%%%%%%%%%%%%%%%%%%%%%%%%%%%%%%%%%%%%%%%%%%%%
\section{Matrix representation for fermionic operators}
\noindent In the second quantization formulation, the behaviour
of a system of fermions on a lattice is described by creation
and annihilation operators, which are governed by the algebra:
\begin{equation}
\{ c^{}_{{i},{s}}, c^{}_{{j},
s'}\} = 0 \hspace{1cm} \{ c^{}_{{i},s} ,
c^{\dagger}_{{j}, s'} \}
= \delta_{{i},{j}}\,
\delta_{s,\,s'}
\label{antic}
\end{equation}
where $s$ is the spin-$J$ label assuming $2J+1$ possible values
($J$ being a half odd integer for fermions).
The local $d$-dimensional vector space $V_j$ associated with
the $j$-th site of the lattice is made up of vectors that are
built by acting with creation operators on a local
vacuum~$| o \rangle_j$:
\begin{equation}
V_j = {\mbox{\it Span}} \left( \, | \alpha \rangle_j = h^{(\alpha)}_j
|o \rangle_j \, , \alpha=1 \ldots d \right)
\label{vectors}
\end{equation}
Here the $h^{(\alpha)}_j $'s are products of $m$ creation operators
$c^{\dagger}_{js}$ with different $s$ ($m$ can
take values from 0 to at most $2J+1$).
We shall tipically deal with spin-$\frac{1}{2}$ fermions, for
which a 4-dimensional space is usually involved, so that
$h^{(1)}_j=c^{\dagger}_{j\,\uparrow}  \, ; \,
h^{(2)}_j=c^{\dagger}_{j\,\downarrow} \, ;
\, h^{(3)}_j=1 \, ; \, h^{(4)}_j = c^{\dagger}_{j\,\downarrow}
c^{\dagger}_{j\,\uparrow}$.
Due to the anticommutation relations (\ref{antic}),
the space $V_j$ has an intrinsic graduation; in fact
$V_j=V^{(1)}_j \oplus V^{(0)}_j$, where the odd (even) subspace
$V^{(1)}_j$ ($\,V^{(0)}_j$) is spanned by those vectors that are built
with an odd (even) number of creation operators~$c^{\dagger}_{i s}$.
Similarly, the space End($V_j$) of local linear operators on
$V_j$ is also graded; odd (even) vectors and
operators are also said to have a parity $p=1$ ($p=0$).
Moreover, since for any homogeneous ${\mathcal{O}}^{(a)}_j ,
{\mathcal{O}}^{(b)}_j \, \in \mbox{End}(V_j)$ the relation
$p({\mathcal{O}}^{(a)}_j \,
{\mathcal{O}}^{(b)}_j)=p({\mathcal{O}}^{(a)}_j)+
p({\mathcal{O}}^{(b)}_j)$ holds,\, End($V$) is actually a graded
local algebra, which we shall denote ${\mathcal{A}}_j$.
Each operator ${\mathcal{O}}^{(a)}_j \in {\mathcal{A}}_j$
can be easily given a {\it local representation} in terms of a
$d$ x $d$ matrix $O^{(a)}$  by making it act on each basis vector
of the space $V_j$
\begin{equation}
{\mathcal{O}}^{(a)}_j \, |\beta \rangle_j  = ({O^{(a)}})^{\alpha}_{\beta} \,
|\alpha \rangle_j
\label{local}
\end{equation}
Here $({O^{(a)}})^{\alpha}_{\beta}$ are ${\Bbb C}$-numbers, $\alpha$
representing the row and $\beta$ the column of the matrix
${O^{(a)}}$. It is easily checked that this representation is
faithful.
\\Let us now consider the fermionic problem on the whole lattice;
let $N$ be the number of sites.
The global algebra $\mathcal{A}$ is the enveloping algebra of the sum of
the local ${\mathcal{A}}_j$'s. Any global operator can be expressed in
terms of linear combination of products of local operators:
\begin{equation}
{\mathcal{O}}_{glob} \in Span \left( {\mathcal{O}}^{(a_{j_1})}_{j_1} \,
{\mathcal{O}}^{(a_{j_2})}_{j_2} \, \ldots
{\mathcal{O}}^{(a_{j_m})}_{j_m}  \right)
\label{O-op}
\end{equation}
where each ${\mathcal{O}}^{(a_{j_i})}_{j_i}$ is a generic
single-site operator belonging to the local sub-algebra
${\mathcal{A}}_{j_i}$ of the
${j}_{i}-$th site, and $m$ can run from 1 to $N$. In general,
the local operators in (\ref{O-op}) need
neither appear in the order of the lattice sites nor
be homogeneous elements of the algebra.
\\It is therefore easily understood that the basic tool
to set up a matrix representation for operators like (\ref{O-op})
is the {\it global representation} of a single-site operator
${\mathcal{O}}^{(a)}_k$: since the latter is embedded in the
global algebra, the matrix that represents it is actually a
multi-index ($d^N$ x $d^N$) matrix, which can be constructed in
principle from the $d$~x~$d$ local representation determining its
action on the local space $V_j$ (see~(\ref{local})). However,
some caution has to be used in doing that, because of the graded
structure of both the operators and the vectors. In particular, a
{\it graded} tensor product must be used instead of an ordinary
one.
\\Therefore, we first of all recall here some notions about multi-index
matrices on graded euclidean vector spaces. A $n$-multi-index
matrix $A$ is a $d^n$ x $d^n$ matrix, whose entries are ${\Bbb
C}$-numbers denoted by $A^{\alpha_1, \ldots \alpha_n}_{\beta_1
\ldots \beta_n }$, each greek index running from 1 to $d$. The
upper $n$-multi-index $\{\alpha \}=(\alpha_1, \ldots \alpha_n)$
represents the rows, while the lower $n$-multi-index $\{\beta \}=
(\beta_1, \ldots \beta_n)$ stands for the columns. The couple
$(\alpha_j, \beta_j)$ describes the action of $A$ on the $j-$th
local ${{\Bbb C}}^d$ space. In each of these euclidean spaces,
the canonical basis vectors $e_{\alpha}$ are supposed to be
assigned a parity $\pi(\alpha)$ \cite{KUL2}, so that ${{\Bbb
C}}^d$ is actually a graded vector space. This function $\pi$ is
in principle completely arbitrary; however we shall choose in the
following $\pi=p$, where $p$ is the {\it intrinsic} parity of the
fermionic vectors $|\alpha\rangle_j$ of $V_j$
(see(\ref{vectors})).
\\The parity of a
$n$-multi-index is defined as
$
p(\{\alpha \}) = p(\alpha_1) + \ldots +p(\alpha_n)
$.
An $n$-multi-index ${\Bbb C}$-number matrix $A$ is
said to be {\it homogeneous} with parity $p(A)$ iff for all
its non-vanishing entires
$A^{ \{ \alpha \} }_{ \{ \beta \}}$ one has
$p(\{\alpha\})+p(\{\beta\})=const=p(A)$.
\\Given a $n$-multi-index matrix $A$ and a $m$-multi-index
matrix $B$ (not necessarily homogeneous), we define their graded
tensor product as
\begin{equation}
\left(A \otimes^{s}_{} B\right)^
{(\{ \alpha \},\{ \gamma \})}_{(\{ \beta \},\{ \delta \})} \,
= \, A^{\{ \alpha \}}_{\{ \beta \}} B^{\{ \gamma \}}_{\{ \delta \}}
\, (-1)^{p(\{ \beta \}) \, (p({\{ \gamma \}})+p({\{ \delta \}}))}
\label{tp^s}
\end{equation}
The graded tensor product is a $(n+m)$-multi-index matrix. If $A$
and $B$ are homogeneous, then $A \otimes^{s}_{} B$ is homogeneous
with parity $p(A)+p(B)$. Although the associativity property $ (A
\otimes^{s}_{} B ) \otimes^{s}_{} C \, = \, A \, \otimes^{s}_{}
\, (B \, \otimes^{s}_{} \, C) $ holds, not any properties of the
ordinary tensor product can be transferred in general to the
graded tensor product; for instance, only if $B$ and $C$ are
homogeneous one can state~that
\begin{equation}
(A {\otimes}^{s}_{} B ) (C {\otimes}^{s}_{} D) =
(-1)^{p(B) p(C)} (AC {\otimes}^{s}_{} BD)
\label{prodotto^s}
\end{equation}
\\Let us now come again to the problem of matrix representation
of fermionic operators. In
order to do that, we remind that for fermions
a convention has to be specified to define the basis
vectors of the global space
$V_{glob}=V \otimes \ldots \otimes V_N$; we shall adopt the
following one
\begin{equation}
| \alpha_1, \alpha_2 ,\ldots \alpha_N \rangle \, \stackrel{def}{=}
\, h^{(\alpha_1)}_1 \ldots  h^{(\alpha_N)}_N | 0 \rangle
\label{bv^s}
\end{equation}
where $| 0 \rangle $ is the global vacuum (defined through
$c^{}_{j s}|0\rangle=0 \; \, \forall \, i,s$), and
$h^{(\alpha_j)}_j$ is the $\alpha_j$-th of the $d$ operators
defining the state vectors at the $j$-th site (see (\ref{vectors})).
The action of any global operator ${\mathcal{O}}_{glob}$ on the
global vectors is perfectly determined by the algebra (\ref{antic}).
Its global representation $O_{glob}$ is defined through:
\begin{equation}
{\mathcal{O}}_{glob}\, |\beta_1, \ldots , \beta_N
\rangle \, = \, (O_{glob})^{\alpha_1 \ldots \alpha_N}_
{\beta_1 \ldots \beta_N} |\alpha_1, \ldots ,\alpha_N
\rangle
\end{equation}
The key step is to provide a matrix representation for single-site
operators; it is straightforwardly realized that, with the choice
(\ref{bv^s}) for the basis vectors, the ${\Bbb C}$-number
$d^N$ x $d^N$ matrix $O^{(a)}_j$ representing ${\mathcal{O}}^{(a)}_j$
as a global operator
turns out to be:
\begin{equation}
O^{(a)}_j = {\Bbb I} \otimes^{s} \ldots \otimes^{s} {\Bbb I}
{\otimes}^{s} \,  \underbrace{O^{(a)}}_{j-\mbox{\tiny th}} \,
{\otimes}^{s} \ldots {\otimes}^{s} {\Bbb I}
\label{repres}
\end{equation}
where $O^{(a)}$ is the $d$ x $d$ matrix representing the
local action of ${\mathcal{O}}^{(a)}_{j}$ on the local
space~(see (\ref{local})). Notice that, thanks to the
associativity property cited above, this definition is
unambiguous.
\\The matrix representation of an operator ${\mathcal{O}}=
{\mathcal{O}}^{(a_{j_1})}_{j_1} \,
{\mathcal{O}}^{(a_{j_2})}_{j_2} \, \ldots
{\mathcal{O}}^{(a_{j_m})}_{j_m}$ is given by the product of the
global representations of the single-site operators
${\mathcal{O}}^{(a_{j_i})}_{j_i}$. In particular, if $j_i = i$
({\it i.e.} the order in which they appear matches the order of
the $h^{(a_j)}_j$'s in the definition (\ref{bv^s}) of the global
vectors), then the matrix is simply given by $ O = O^{(a_1)}_1
{\otimes}^{s} \ldots {\otimes}^{s} O^{(a_m)}_m $. We wish to
point out that the matrix representation introduced here also
holds for non-homogeneous operators: this is very useful because
one is not required to decompose a given operator into its even
and odd
components before arriving at its matrix representation.\\

A crucial role will be played in the following by Hubbard
projectors ${{\mathcal{E}}_{j}}^{b}_{a}=|a\rangle_j \,
{}_j\langle b|$. In the case of a 4-dim. local space they are
explicitly given by the entries ($a-$th row and the $b-$th column)
of the following matrix\\
\begin{equation}
{\mathcal{E}}_{j} \, = \, \pmatrix{
n^{}_{j\uparrow}(1-n^{}_{j\downarrow}) &
c^{\dagger}_{j\uparrow} c^{}_{j\downarrow} &
c^{\dagger}_{j\uparrow} (1-n^{}_{j\downarrow}) &
c^{}_{j\downarrow} n^{}_{j\uparrow} \cr
%%%%
c^{\dagger}_{j\downarrow} c^{}_{j\uparrow} &
n^{}_{j\downarrow} (1-n^{}_{j\uparrow}) &
c^{\dagger}_{j\downarrow} (1-n^{}_{j\uparrow}) &
-c^{\dagger}_{j\uparrow} n^{}_{j\downarrow} \cr
%%%%
c^{}_{j\uparrow} (1-n^{}_{j\downarrow}) &
c^{}_{j\downarrow} (1-n^{}_{j\uparrow}) &
(1-n^{}_{j\uparrow})(1-n^{}_{j\downarrow}) &
c^{}_{j\uparrow} c^{}_{j\downarrow} \cr
%%%%
c^{\dagger}_{j\downarrow} n^{}_{j\uparrow} &
-c^{}_{j\uparrow} n^{}_{j\downarrow} &
c^{\dagger}_{j\downarrow} c^{\dagger}_{j\uparrow} &
n^{}_{j\downarrow} n^{}_{j\uparrow} \cr
}
\label{Hubb-proj}
\end{equation}
\\Each of the above entries is an homogeneous operator with parity
$
p({{\mathcal{E}}_j}^{b}_{a}) \, = \, p(a)+p(b)
$.
The Hubbard projectors enjoy very important properties:
\begin{eqnarray}
\left[
{{\mathcal{E}}_{j}}^{b}_{a} \, , \, {{\mathcal{E}}_{k}}^{d}_{c}
\right]_{\pm} = 0 \hspace{1cm} \forall j \neq k
\label{prop-1}
\\
{{\mathcal{E}}_{j}}^{b}_{a} \, {{\mathcal{E}}_{j}}^{d}_{c} =
\delta^{b}_{c} \, {{\mathcal{E}}_{j}}^{d}_{a} \hspace{2.2cm}
\label{prop-2}
\end{eqnarray}
where $[ X , Y]_\pm=X \, Y -(-1)^{p(X)p(Y)} \, Y X$.
The matrix representation of the Hubbard projectors will be
denoted ${E_j}^{b}_{a}$; it is given by eqn.(\ref{repres}) where
${O}^{(a_j)} \rightarrow {E_{}}^{b}_{a}$, the $d \times d$
matrix ${E_{}}^{b}_{a}$ having vanishing entries except for a 1 at
row $a$ and column $b$.
The ${E_j}^{b}_{a}$'s share the same properties (\ref{prop-1}) and
(\ref{prop-2}) as the ${{\mathcal{E}}_{j}}^{b}_{b}$.
Using the Hubbard projectors, one can express any single-site operator
${\mathcal{O}}^{(a)}_j$ just using its {\it local} representing matrix
(see (\ref{local})) as follows
\begin{equation}
{\mathcal{O}}^{(a)}_j=({O}^{(a)})^{\alpha}_{\beta} \,
{{\mathcal{E}}_{j}}^{\beta}_{\alpha}
\label{local-local}
\end{equation}
\label{sec-2}
%%%%%%%%%%%%%%%%%%%%%%%%%%%%%%%%%%%%%%%%%%%%%%%%%%%%%%%%%%%%%%%%
%%%%%%%%%%%%%%%%%%%%%%%%%%%%%%%%%%%%%%%%%%%%%%%%%%%%%%%%%%%%%%%%
\section{The Quantum Inverse Scattering Method for fermionic systems}
\noindent
The Quantum Inverse Scattering Method (QISM) is a powerful tool
for studying quantum integrability because it provides a set of
mutually commuting operators. Within the QISM
a key role is played by the so called  ${\mathcal{L}}$-operator,
an operator-valued matrix acting on an
$n$-dimensional space termed {\it auxiliary}; the
${\mathcal{L}}$-operator is local, so that in discrete 1-dimensional systems
one has a matrix ${\mathcal{L}}_j$ for each site $j$ of a chain. The
nature of its entries ${{\mathcal{L}}_j}^{\alpha}_{\beta}$ determines
the kind of physical model one is dealing with ({\it i.e.} the ${{\mathcal{L}}_j}^{\alpha}_{\beta}$ are spin operators for spin
systems, fermionic operators for fermionic models, etc.); they
also depend on two complex parameters which are referred to as
spectral parameters $u$ and $v$. When the
${{\mathcal{L}}_j}^{\alpha}_{\beta}$ belong to an ordinary algebra
${{\mathcal{G}}_j}$, the standard formulation of QISM can be applied
as follows: if the commutation rules of the ${\mathcal{L}}$-operator
can be expressed in terms of a $n^2 \times n^2$ ${\Bbb C}$-number
matrix $\Rc(u,v)$ through the relation
$
\Rc \, ({{\mathcal{L}}_j} \otimes^{}
{{\mathcal{L}}_j} ) \, = \, ({{\mathcal{L}}_j}
\otimes^{} {{\mathcal{L}}_j} ) \, \Rc
$ (local realization of the Yang-Baxter algebra),
then a global realization
$
\Rc \, ({{\mathcal{T}}} \otimes^{}
{{\mathcal{T}}} ) \, = \, ({{\mathcal{T}}}
\otimes^{} {{\mathcal{T}}} ) \, \Rc
$ can be deduced, where ${\mathcal{T}}_{N}={\mathcal{L}}_{N} \ldots
{\mathcal{L}}_{1}$ is the monodromy matrix. From this property
a set of mutually commuting operators can be obtained\cite{KUL,KBI}.
The crucial step to pass from the local to the global realization is the
relation
$
({{\mathcal{L}}_i} \otimes^{} {{\mathcal{L}}_j} )
({{\mathcal{L}}_k} {\otimes}^{} {{\mathcal{L}}_l}) =
({{\mathcal{L}}_i} {{\mathcal{L}}_k} {\otimes}^{}
{{\mathcal{L}}_j}  {{\mathcal{L}}_l}) \, \,
\forall \, j \neq k
$, which stems from the fact that for ordinary algebras the local
operators referring to different sites commute.
\\However, when dealing with fermionic systems, the entries
of the ${\mathcal{L}}$-operator belong to a superalgebra
${\mathcal{G}}^{s}_{j}$, ${{\mathcal{L}}_j}^{\alpha}_{\beta}$ and
${{\mathcal{L}}_k}^{\gamma}_{\delta}$ may
not commute, even when involving different sites $j$ and $k$.
Therefore, in order to work out a QISM for
fermionic systems, a different composition law (other than $\otimes$)
has to be introduced.
To this purpose, it is customary to assume that the
${\mathcal{L}}$-operator is homogeneous.
We recall that in general a matrix $A$ with Grassmann valued entries
is said to be homogeneous with parity $p(A)$ iff for all its
non-vanishing entries
$p(A^{ \{\alpha\}}_{\{\beta\}})+p(\{\alpha\})+p(\{\beta\})=const=p(A)$,
where $p(A^{ \{\alpha\}}_{\{\beta\}})$ is the Grassmann parity of the
entry at multi-row $\{ \alpha \}$ and multi-column $\{ \beta \}$
(notice that for ${\Bbb C}$-number matrices one has
$p(A^{ \{\alpha\}}_{\{\beta\}})=0$ and one recovers the definition
given in section II).
\\In practice one usually deals with even ${\mathcal{L}}$-operators
({\it i.e.} $p({\mathcal{L}})=0$); as a consequence ${\mathcal{T}}$
is even too; in this case the required composition law is given by a
graded tensor product defined as follows
\begin{equation}
\left({\mathcal{A}} {\otimes}^{}_{s} {\mathcal{B}}\right)^
{(\{ \alpha \},\{ \gamma \})}_{(\{ \beta \},\{ \delta \})} \, = \,
{\mathcal{A}}^{\{ \alpha \}}_{\{ \beta \}} \,
{\mathcal{B}}^{\{ \gamma \}}_{\{ \delta \}} \,
(-1)^{(p(\{\alpha\})+p(\{\beta\})) \, p(\{\gamma\})}
\label{tp_s}
\end{equation}
for arbitrary ${\mathcal{G}}^s$-valued matrices $\mathcal{A}$
and $\mathcal{B}$ (where ${\mathcal{G}}^s$ is the enveloping
algebra of the locals ${\mathcal{G}}_{j}^s \,$'s).
It is possible to check that $\otimes_s$ is associative and that
\begin{equation}
({{\mathcal{L}}_i} {\otimes}^{}_{s} {{\mathcal{L}}_j} )
({{\mathcal{L}}_k} {\otimes}^{}_{s} {{\mathcal{L}}_l}) =
({{\mathcal{L}}_i} {{\mathcal{L}}_k} {\otimes}^{}_{s}
{{\mathcal{L}}_j}  {{\mathcal{L}}_l}) \hspace{1cm} \forall \, j \neq k
\label{prodotto-per-L}
\end{equation}
as it is expected to be for even objects. Thanks to (\ref{prodotto-per-L}),
one can state\cite{OWA} that
\begin{eqnarray}
\Rc(u,v) \, ({{\mathcal{L}}}(u,w) \otimes_{s}
{{\mathcal{L}}}(v,w) ) \, &=& \, ({{\mathcal{L}}}(v,w)
\otimes_{s} {{\mathcal{L}}}(u,v) ) \, \Rc(u,v)\label{RLLg} \\
\hspace{1cm} & \Downarrow & \hspace{1cm} \nonumber \\
\Rc(u,v) \, ({{\mathcal{T}}}(u,w) \otimes_{s}
{{\mathcal{T}}}(v,w) ) \, &=& \, ({{\mathcal{T}}}(v,w)
\otimes_{s} {{\mathcal{T}}}(u,v) ) \, \Rc(u,v) \quad \quad .
\label{RTTg}
\end{eqnarray}
The fact that for fermionic $\mathcal{L}$-operators a graded
tensor product has to be introduced also leads to assuming
that the $\Rc$-matrix is even too, {\it i.e.}
\begin{equation}
p(\alpha)+p(\beta)+p(\gamma)+p(\delta)=0 \hspace{1cm} \forall \,
\Rc^{\alpha \gamma}_{\beta\delta}(u,v) \neq 0 \quad \quad .
\label{pari}
\end{equation}
Indeed, this turns out to be very natural when investigating the
consistency conditions of eqn.(\ref{RLLg}); in fact, one can
mimic the non-graded case \cite{KBI} and try to derive the
Jacobi identities for the structure constants $\Rc$ by reducing
a multiple product ${{\mathcal{L}}_j}^{\prime} \otimes_s
{{\mathcal{L}}_j}^{\prime \prime} \otimes_s
{{\mathcal{L}}_j}^{\prime \prime \prime}$ into its `reversed' form
${{\mathcal{L}}_j}^{\prime \prime \prime} \otimes_s
{{\mathcal{L}}_j}^{\prime \prime} \otimes_s
{{\mathcal{L}}_j}^{\prime}$ through repeated action of eqn.(\ref{RLLg}).
A compact form is obtained only if $\Rc$ is even too;
moreover it can be shown that, despite the fact that graded algebras are
involved, the obtained relation is {\it non-}graded
\begin{equation}
({\Bbb I} \otimes \Rc(u,v)) \, (\Rc(u,w) \otimes {\Bbb I}) \,
({\Bbb I} \otimes \Rc(v,w)) \, = \,
(\Rc(v,w) \otimes {\Bbb I}) \, ({\Bbb I} \otimes \Rc(u,w)) \,
(\Rc(u,v) \otimes {\Bbb I}) \label{YBE-Rc-comp}
\end{equation}
This ${\Bbb C}$-number functional equation is known as the
ordinary (because no extra signs appear) Yang-Baxter Equation
(YBE). It is worth stressing that $\Rc$ may depend on the
spectral parameters in a completely general way, not only through
their difference.
\\In analogy to what happens for spin models, the above scheme
can also be approached backward, {\it i.e.} the YBE itself can be
used to explicitly realize the structure (\ref{RLLg}), then
yielding a set of mutually commuting fermionic operators. A
peculiar feature of this procedure is that the dimension $n$ of
the auxiliary space is taken to be equal to the dimension $d$ of
the physical local space. Interestingly, one can prove
\cite{GOMU,KOR} that, starting from an even (see
eqn.(\ref{pari})) solution of the YBE (\ref{YBE-R-comp}), one can
construct an even $\mathcal{L}$-operator as follows
\begin{equation}
{{\mathcal{L}}_j}^{\alpha}_{\beta}(u,v)=(-1)^{p(\alpha) p(\gamma)}
\Rc^{\gamma \alpha}_{\beta\delta}(u,v) \,
{{\mathcal{E}}_j}^{\delta}_{\gamma}
\label{L-op}
\end{equation}
where the ${{\mathcal{E}}_j}^{\delta}_{\gamma}$ are the Hubbard
projectors. Here $\alpha$ gives the row and $\beta$ the column
of the entry. Due to eqn.(\ref{prop-1}) the entries of two operators
${\mathcal{L}}_j$
and ${\mathcal{L}}_k$ of different sites fulfill
\begin{equation}
\left[{{\mathcal{L}}_j}^{\alpha}_{\beta}(u,v),{{\mathcal{L}}_k}^
{\gamma}_{\delta}(u',v')\right]_{\pm}=0 \hspace{0.3cm}
\forall \alpha,\beta,\gamma,\delta
\hspace{0.5cm} \forall u,v,u',v' \hspace{0.5cm} \forall j\neq k
\label{L-op-sc}
\end{equation}
At the same time the property (\ref{prop-2}) allows to show
that the Yang-Baxter equation (\ref{YBE-Rc-comp}) is actually
equivalent to (\ref{RLLg}). In doing so, one can easily realize
that the presence of the additional signs in the definition
(\ref{L-op}) of the ${\mathcal{L}}$-operator is crucial; therefore
in the graded case, it is also customary to rewrite
eqn.(\ref{YBE-Rc-comp}) for the variables
$\tilde{R}^{\alpha \gamma}_{\beta \delta}=(-1)^{p(\alpha) p(\gamma)}
\Rc^{\gamma \alpha}_{\beta \delta}$, obtaining the so-called
graded Yang-Baxter equation\cite{KUL,KUL2}
\begin{eqnarray}
\lefteqn{\tilde{R}^{\alpha\beta}_{\alpha'\beta'}(u,v) \,
\tilde{R}^{\alpha'\gamma}_{\alpha''\gamma'}(u,w) \,
\tilde{R}^{\beta'\gamma'}_{\beta''\gamma''}(v,w) (-1)^{p(\beta')
(p(\alpha') + p(\alpha''))} \, = \hspace{1.5cm} \label{gYBE}}  \\
& & \tilde{R}^{\beta\gamma}_{\beta'\gamma'}(v,w) \,
\tilde{R}^{\alpha\gamma'}_{\alpha'\gamma''}(u,w) \,
\tilde{R}^{\alpha'\beta'}_{\alpha''\beta''}(u,v) \,(-1)^{p(\beta')
(p(\alpha) + p(\alpha'))} \quad.
\nonumber
\end{eqnarray}
Starting from eqn.(\ref{RTTg}) one can derive a set of
commuting fermionic operators. It is worth emphasizing that,
since $\Rc(u,v)$ is an even matrix, two
relations are obtained
\begin{equation}
\left[ tr \, {{\mathcal{T}}_N}(u,w) ,  tr \,
{{\mathcal{T}}_N}(v,w) \right] = 0 \hspace{2cm}
\left[ s tr \, {{\mathcal{T}}_N}(u,w) ,  str \,
{{\mathcal{T}}_N}(v,w) \right] = 0 \label{cons-tr-str}
\end{equation}
where $tr$ is the ordinary trace and $str$ is the supertrace
$str\, {\mathcal{T}}=(-1)^{p(\alpha)}
{\mathcal{T}}^{\alpha}_{\alpha}$. Therefore it might seem that
one has two kinds of conservation laws; however
it is not by developing the equations (\ref{cons-tr-str})
powers of the spectral parameters that one generates
appropriate constants of motion, because neither
of them yields local operators; in order to have such a property
one usually requires to have a couple
of values $(u_0,v_0)$ of the spectral parameters for which the
solution of eqn.(\ref{YBE-Rc-comp}) reduces
\begin{equation}
\Rc^{\alpha \gamma}_{\beta\delta}(u_0,v_0)=\delta^{\alpha}_{\beta}
\, \delta^{\gamma}_{\delta}
\label{cond-2}
\end{equation}
If this is the case, a straightforward calculation shows that:
\begin{equation}
{\mathcal{Z}}(u,v) := (str \, {\mathcal{T}}_N (u_0,v_0))^{-1}
\, str \, {\mathcal{T}}_N (u,v) \, \equiv \,
(tr \, {\mathcal{T}}_N (u_0,v_0))^{-1} \, tr \,
{\mathcal{T}}_N (u,v) \label{Z}
\end{equation}
and that the constants of motion defined as
\begin{equation}
{\mathcal{J}}_n \, = \, \left. \frac{d^n}{d u^n} \ln
{\mathcal{Z}}(u,v_0) \right|_{u=u_0}
\hspace{1cm} n \ge 1 \label{J_n} \quad .
\end{equation}
are local, in that ${\mathcal{J}}_n$ is the sum of
operators involving clusters of no
more than $n+1$ sites\cite{KUL}.
The logarithm is taken not only to obtain an additive eigenvalue
spectrum in performing the Algebraic Bethe Ansatz\cite{FAD,RAMA,RAMA3},
but also to avoid trivially
commuting constants.
Eqn.(\ref{Z}) shows that, although the transfer matrix is usually
defined through the supertrace
\begin{equation}
\tau (u,v)=str \, {\mathcal{T}}_N (u,v) \label{tau}
\end{equation}
the ordinary trace plays an equivalent role. However, taking the
definition (\ref{tau}) does simplify the expression of the
shift-operator, because $str \, {\mathcal{T}}_N (u_0,v_0)=
{\mathcal{P}}^{g}_{12}{\mathcal{P}}^{g}_{23} \ldots
{\mathcal{P}}^{g}_{N-1\,N}$,
where ${\mathcal{P}}^{g}_{jk}$ is in this case the {\it graded}
permutator
$
{\mathcal{P}}^{g}_{jk}=(-1)^{\beta}
{{\mathcal{E}}_j}^{\beta}_{\alpha} \,
{{\mathcal{E}}_k}^{\alpha}_{\beta}
$.
Such a simple and compact form is not possible for the
$tr \, {\mathcal{T}}_N (u,v) $.
A deep argument confirming the crucial role of the
super-trace will be provided by the investigation of the
symmetries (see section~VI).
\\Let us consider in detail the first conserved quantity, usually
interpreted as the Hamiltonian, which is of the form
$
{\mathcal{H}}=\sum_{j=1}^{N} {\mathcal{H}}_{j \, j+1}
$,
with periodic boundary conditions ${\mathcal{H}}_{N \, N+1}=
{\mathcal{H}}_{N \,1}$. Each ${\mathcal{H}}_{j \, j+1}$ is a
two-site Hamiltonian given by
\begin{equation}
{\mathcal{H}}_{j \, j+1}=(-1)^{p(\gamma) (p(\beta)+p(\delta))} \left.
\partial_u \check{R}^{\alpha \beta}_{\gamma \delta}(u,v_0) \right|_{u=u_0}
{{\mathcal{E}}_j}^{\gamma}_{\alpha} \,
{{\mathcal{E}}_{j+1}}^{\delta}_{\beta}
\label{2-site-Ham-op}
\end{equation}
Writing down the matrix representation for the Hubbard projectors,
it easily verified that the matrix representation
$H_{j \, j+1}$ for the two-site Hamiltonian (\ref{2-site-Ham-op})
is
\begin{equation}
{H}_{j \, j+1} = {\Bbb I} {\otimes}^{s} \ldots
{\otimes}^{s} \,  \underbrace{H_{\mbox{\tiny 2 sites}}}_{j\, j+1} \,
{\otimes}^{s} \ldots {\otimes}^{s} {\Bbb I} \hspace{2cm}
(H_{2\,\mbox{\tiny sites}})^{\alpha \beta}_{\gamma \delta}= \left.
\partial_u \check{R}^{\alpha \beta}_{\gamma \delta}(u,v_0) \right|_{u=u_0}
\label{2-site-Ham-mat2}
\end{equation}
Finally, we also remark that the fermionic operators
$\mathcal{R}$ and $\check{{\mathcal{R}}}$ can be
constructed as
\begin{eqnarray}
{\check{\mathcal{R}}}_{j \, k} &=& (-1)^{p(\gamma)
(p(\beta)+p(\delta))}
\check{R}^{\alpha \beta}_{\gamma \delta}(u,v) \,
{{\mathcal{E}}_j}^{\gamma}_{\alpha} \,
{{\mathcal{E}}_{k}}^{\delta}_{\beta}
\label{ferm-Rc-matr} \\
{\mathcal{R}}_{j \, k} &=& (-1)^{p(\gamma) (p(\beta)+p(\delta))}
\tilde{R}^{\alpha \beta}_{\gamma \delta}(u,v) \,
{{\mathcal{E}}_j}^{\gamma}_{\alpha} \,
{{\mathcal{E}}_{k}}^{\delta}_{\beta} \, = \,
{\mathcal{P}}^{g}_{j \, k} \,
\check{{\mathcal{R}}}_{j \, k}
\label{ferm-R-matr}
\end{eqnarray}
Using $\check{{\mathcal{R}}}_{j \, k}$ and ${\mathcal{R}}_{j \, k}$,
one can rewrite eqns.(\ref{YBE-Rc-comp}) and
(\ref{gYBE}) in equivalent forms
\begin{eqnarray}
\check{{\mathcal{R}}}_{23}(u,v) \, \check{{\mathcal{R}}}_{12}(u,w) \, \check{{\mathcal{R}}}_{23}(v,w) \, = \,
\check{{\mathcal{R}}}_{12}(v,w) \, \check{{\mathcal{R}}}_{23}(u,w) \, \check{{\mathcal{R}}}_{12}(u,v) \label{YBE-Rc-ferm}  \\
{\mathcal{R}}_{12}(u,v) \, {\mathcal{R}}_{13} (u,w) \,
{\mathcal{R}}_{23}(v,w)\, = \,
{\mathcal{R}}_{23}(v,w) \, {\mathcal{R}}_{13} (u,w)\,
{\mathcal{R}}_{12}(u,v)
\end{eqnarray}
Also, eqn.(\ref{2-site-Ham-op})
can be unified into the shorter form
\begin{equation}
{\mathcal{H}}_{j \, j+1}= \left.
\partial_u \check{{\mathcal{R}}}^{\alpha \beta}_
{\gamma \delta}(u,v_0) \right|_{u=u_0}
\label{Ham-from-Rc-op}
\end{equation}
\label{sec-3}
%%%%%%%%%%%%%%%%%%%%%%%%%%%%%%%%%%%%%%%%%%%%%%%%%%%%%%%
%%%%%%%%%%%%%%%%%%%%%%%%%%%%%%%%%%%%%%%%%%%%%%%%%%%%%%%
\section{Some remarks on the graded tensor product}
\noindent Throughout the previous sections we used two different
kinds of graded tensor product~({\it gtp}); the former was
introduced to give matrix representations of fermionic operator
({\it REP-gtp}), the latter to adapt the QISM to fermionic
systems ({\it QISM-gtp}). We defined and denoted them differently
(${\otimes}^{s}$ and ${\otimes}_{s}$ respectively) to emphasize
their different roles. Indeed, in order to explicitly convert
fermionic operator into matrices or to correctly interpret matrix
results in terms of fermionic operators, it is worth
distinguishing them; since in the literature they are often
exchanged and the conventions yielding them are not always
precised, we wish to comment about that. The graded tensor product
$\otimes^s$ (see (\ref{tp^s})) fits to represent the fermionic
operators when the convention on the definition of the basis
vectors is (\ref{bv^s}); however, had we chosen $ | a_1, a_2
,\ldots a_N \rangle \, \stackrel{def}{=} \, h^{(a_N)}_N \ldots
h^{(a_1)}_1 | 0 \rangle $, we would have got a product like
${\otimes}^{}_{s}$ (see (\ref{tp_s})). The {\it REP-gtp} is
therefore strictly related to the convention adopted on the basis
vectors.
\\On the contrary, the {\it QISM-gtp} has different origins: it must
be introduced in order to obtain eqn.(\ref{prodotto-per-L}), which
could not be derived from an ordinary tensor product. The actual
explicit form ${\otimes}_s$ used for the {\it QISM-gtp} just stems
from the additional signs in the definition (\ref{L-op}) of the
$\mathcal{L}$-operator; there the indices $\alpha$ and $\gamma$
are the {\it rows} of $\Rc$. This choice is customary
in the literature; however, one could equivalently
introduce such extra signs in correspondence with the columns,
obtaining that the YBE~(\ref{YBE-Rc-comp}) is
equivalent to an expression similar to (\ref{RLLg}) where
${\otimes}_{s}$ is replaced by ${\otimes}^{s}$.
\\Thus, one could in principle use either
${\otimes}_{s}$ or ${\otimes}^{s}$
to define both {\it REP-gtp} and {\it QISM-gtp}; however, the
use of two {\it different} definitions for the two
graded tensor products leads to useful simplifications
in the interpretation of the matrix results.
In fact, the expression (\ref{2-site-Ham-op}) for the two-site
Hamiltonian ${\mathcal{H}}_{j \, j+1}$ contains some additional signs,
which stem from the definition~(\ref{L-op}) of the $\mathcal{L}$-operator
(so, they are related to the choice of ${\otimes}^{}_{s}$ for
{\it QISM-gtp}); as noticed in sec.III, when writing down the
matrix representation ${\mathcal{H}}_{j\,j+1}$, the above additional
signs cancel out with the signs coming from the matrix representation
of the Hubbard projectors, provided that  ${\otimes}^{s}$ is used as
{\it REP-gtp}. As a consequence, eqn.(\ref{2-site-Ham-mat2}) contains
no more extra signs. On the contrary, if the {\it same} definition
${\otimes}^{}_{s}$ was adopted for both {\it REP-gtp} and {\it QISM-gtp},
further additional signs would add to the previous ones, yielding:
\begin{equation}
\left(H_{2\,\mbox{\tiny sites}}\right)^{\alpha \gamma}_{\beta \delta} =
(-1)^{(p(\alpha)+p(\gamma))(p(\beta) +p(\delta))}\left.
\partial_u \check{R}^{\alpha \gamma}_{\beta \delta}(u,v_0) \right|_{u=u_0}
\end{equation}
which would contain `undesirable' extra signs.
\label{sec-4}
%%%%%%%%%%%%%%%%%%%%%%%%%%%%%%%%%%%%%%%%%%%%%%%%%%%%%%%%
\section{The invariance properties of the YBE}
The invariance transformations of a given equation are those
transformations that map solutions of the equation into other
solutions of the same equation. In the case of the YBE the
invariance properties are important to understand if two models
are {\it independently} integrable or not. A very important
example is supplied by the similarity transformations: if a
system ${\mathcal{H}}$ is integrable and we transform the
constants of motions through a similarity operator
${\mathcal{A}}_{glob}$ as follows ${\mathcal{J}}_n \rightarrow
{\mathcal{J}}^{\prime}_n= {\mathcal{A}}_{glob} \, {\mathcal{J}}_n
\, {\mathcal{A}}^{-1}_{glob}$, the new operators yield a further
integrable system. Nevertheless, it is not that obvious in
general that the new system can be derived from an
$R$-matrix\footnote{for instance, the transformed Hamiltonian may
not be of the form ${\mathcal{H}}^{\prime} =\sum_{i=1}^{N}
{\mathcal{H}}^{\prime}_{i \, i+1}$} because local structures may
not be conserved; however, this is expected to be the case when
the transformation is the product of the same local operator,
{\it i.e.} ${\mathcal{A}}_{glob}={\mathcal{A}}_1 \ldots
{\mathcal{A}}_{N}$, because one is reconducted to study the local
change ${\mathcal{H}}_{j,j+1} \rightarrow ({\mathcal{A}}_j
{\mathcal{A}}_{j+1}) {\mathcal{H}}_{j,j+1}
({\mathcal{A}}^{-1}_{j+1} {\mathcal{A}}^{-1}_{j})$. Actually, for
the non-graded case, Kulish and Sklyanin\cite{KUL} proved that
this result can still be cast in the language of QISM by noticing
that the eq.(\ref{YBE-Rc-comp}) is invariant under similarity
transformation of the form $A \otimes A$; explicitly, if a $n^2
\times n^2$ matrix $\Rc$ is a solution of (\ref{YBE-Rc-comp}) and
$A$ is a $n \times n$ invertible matrix, then the matrix
$\Rc^{\prime}=(A \otimes A) \, \Rc (A \otimes A)^{-1}$ is also a
solution of (\ref{YBE-Rc-comp}).
\\However, as to fermionic systems, such a kind of
similarity transformations are not suitable; in fact the matrix
$A$ still represents a local operator ${\mathcal{A}}_j$, but the
matrices representing ${\mathcal{A}}_j {\mathcal{A}}_{j+1}$ are
built up with the {\it graded} tensor product, {\i.e.} are of the
form $A \otimes^{s} A$; we shall refer to them as graded
similarity transformations. The question therefore arises whether
the YBE equation is actually invariant under such kind of
transformations. To this purpose some facts must be taken into
account:
\\i)
In section III we pointed out that the Yang-Baxter Equations
(\ref{YBE-Rc-comp}) can be set into other forms: we observed that
when using the variable $\tilde{R}=P^g \, \Rc$ --~ where $P^g$ is
the graded permutator $(P^g)^{\alpha \gamma}_{\beta \delta}=
(-1)^{p(\alpha) p(\gamma)} \delta^{\alpha}_{\delta} \,
\delta^{\gamma}_{\beta}$~-- the equation (\ref{gYBE}) is obtained;
in non-graded systems a widely used form is achieved rewriting
(\ref{YBE-Rc-comp}) in terms of $R=P \, \Rc$, where $P$ is the
ordinary permutator $(P^g)^{\alpha \gamma}_{\beta \delta}=
\delta^{\alpha}_{\delta} \, \delta^{\gamma}_{\beta}$, arriving~at
\begin{equation}
R_{12}(u,v) \, R_{13}(u,w) \, R_{23} (v,w) \, = \, R_{23}(v,w) \,
R_{13}(u,w) \, R_{12} (u,v) \label{YBE-R-comp} \quad.
\end{equation}
Although all the above cited forms are equivalent (as long as the
matrices are even), it is not trivial in general that if the
equation in a given form fulfills a certain invariance property,
the equation obtained with a change of variable fulfills the {\it
same} invariance property. Now, it turns out that in the
non-graded case the invariance property under similarity
transformation $A \otimes A$ is actually satisfied from both
(\ref{YBE-Rc-comp}) and (\ref{YBE-R-comp}); in fact the passages
$\Rc \rightarrow \Rc^{\prime} \rightarrow {R}^{\prime}$ and $\Rc
\rightarrow {R} \rightarrow {R}^{\prime}$ yield unambiguously the
same result, since $P (A \otimes A) P=A \otimes A$ for any matrix
$A$. On the contrary, in the graded case the passage $\Rc
\rightarrow \Rc^{\prime} \rightarrow \tilde{R}^{\prime}$ has no
more the same effect as $\Rc \rightarrow \tilde{R} \rightarrow
\tilde{R}^{\prime}$, because in general $P^g (A \otimes^{s} A)
P^g \neq A \otimes^s A$. Thus one has to decide which equation
(either (\ref{YBE-Rc-comp}) or (\ref{YBE-R-comp})) is worth being
envisaged. Since the Hamiltonian is the derivative of $\Rc$ and
not of $R$ (see eqn.(\ref{2-site-Ham-mat2})),
eqn.(\ref{YBE-Rc-comp}) seems to be more appropriate.
\\ii) the even parity of the $R$-matrix (which is
crucial for constructing integrable fermionic systems from the
YBE), may not be conserved under such a similarity transformation.
Therefore, in order for the transformed matrix $\Rc^{\prime}$ to
actually generate a {\it fermionic} system, we must take care
that $p(\Rc^{\prime})=0$. Interestingly, we shall show that such
a (physical) requirement is actually the only one that is
necessary to prove that the (mathematical) invariance property
holds. A simple condition to ensure $p(\Rc^{\prime})=0$ is to
take a homogeneous $A$, because in that case $A \otimes^{s} A$ is
always even (regardless of the parity of $A$) and $\Rc^\prime$ is
the product of 3 even matrices. Nevertheless, other looser
conditions are also applicable in principle, depending on both
the structure of $R$ and the local $A$ to be considered.
\\We can formulate our result as follows: let $A$ be an invertible
$n \times n$ matrix and $\Rc$ an even solution of
(\ref{YBE-Rc-comp}); if the matrix $ \Rc^{\prime}(u,v)=(A
\otimes^s A) \, \Rc(u,v) (A \otimes^s A)^{-1} $ is also even,
then $\Rc^{\prime}$ is still a solution of (\ref{YBE-Rc-comp}).
\\{\it \noindent Proof:} First of all, it is easily realized
that, due to the fact that $\Rc$ is even, the ordinary tensor
product in (\ref{YBE-Rc-comp}) can be replaced by the graded
tensor product $\otimes^{s}$. Then, the following identities are
the key for the proof
\begin{eqnarray}
(A \otimes^{s} A \otimes^{s} A) ({\Bbb I} \otimes^s \Rc) \, =
\, ({\Bbb I} \otimes^s \Rc^{\prime}) \, (A \otimes^{s} A \otimes^{s} A)\\
(A \otimes^{s} A \otimes^{s} A) (\Rc \otimes^{s} {\Bbb I}) \, =
\, (\Rc^{\prime} \otimes^{s} {\Bbb I}) \, (A \otimes^{s} A
\otimes^{s} A)
\end{eqnarray}
In order to prove them, one can multiply both equations on the
left by $(A \otimes^s A \otimes^s A)^{-1}\,(A \otimes^s A
\otimes^s A)$, and realize that, although for the graded tensor
product $(A \otimes^s A \otimes^s A)^{-1} \neq (A^{-1} \otimes^s
A^{-1} \otimes^s A^{-1})$, it is always possible to write
\[(A \otimes^s A \otimes^s A)^{-1}=({\Bbb I} \otimes^s
(A \otimes^s A)^{-1}) (A^{-1} \otimes {\Bbb I} \otimes^{s} {\Bbb
I}) =({\Bbb I} \otimes^s {\Bbb I} \otimes^s A^{-1}) ((A \otimes^s
A)^{-1} \otimes^{s} {\Bbb I})\] Moreover, from the property
(\ref{prodotto^s}), one can obtain that $(A {\otimes}^{s}_{} B )
(C {\otimes}^{s}_{} D) = (AC {\otimes}^{s}_{} BD) $, provided
that $B$ (or $C$) is even, also if the remaining 3 matrices are
not homogeneous. Finally, multiplying eqn.(\ref{YBE-Rc-comp}) by
$A \otimes^{s} A \otimes^{s} A$ on the left, one obtains that
$\Rc^{\prime}$ also fulfills~eqn.(\ref{YBE-Rc-comp}).
\label{sec-5}
%%%%%%%%%%%%%%%%%%%%%%%%%%%%%%%%%%%%%%%%%%%%%%%%%%%%%%%%
%%%%%%%%%%%%%%%%%%%%%%%%%%%%%%%%%%%%%%%%%%%%%%%%%%%%%%%%
%%%%%%%%%%%%%%%%%%%%%%%%%%%%%%%%%%%%%%%%%%%%%%%%%%%%%%%%
%%%%%%%%%%%%%%%%%%%%%%%%%%%%%%%%%%%%%%%%%%%%%%%%%%%%%%%%
\section{\boldmath The symmetries of the constants of motion}
\noindent In the literature, quite powerful algebraic methods
have been developed to classify the solutions of the YBE
(\ref{YBE-Rc-comp}) according to several kinds of symmetries and
supersymmetries. An important question is concerned with effects
that these local symmetries have on the constants of motion
${\mathcal J}_n$ derived from the QISM (eq.(\ref{J_n})). In the
literature this issue has been examined for some specific
fermionic models, such as the $t-J$\cite{FOKA} and the
EKS\cite{EKS4}. The standard approach for these models is to
perform an asymptotic expansion in the (additive) spectral
parameter $u-v \rightarrow \infty$ of the monodromy matrix
${\mathcal T}$, and to use the $RTT=TTR$ eq.(\ref{RTTg}) to
extract from it the generators of the related superalgebra
($spl(2,1)$ for the $t-J$ and $u(2|2)$ for the EKS). The
${\mathcal J}_n$ of these particular cases can thus be proved to
share the symmetries of $\Rc$.
\\In the authors' opinion, however, no sufficiently general
approach has been devoted to this subject. To this purpose, in
this section, by only making use of the existence of the
shift-point condition (\ref{cond-2}), we shall prove that the
local constraints imposed on the $\Rc$-matrix do reflect onto the
whole set of constants of motion. We wish to stress that, with
respect to the known results concerning specific models cited
above, our theorem is more general in that: i) it is independent
of the dimension of the $R$-matrix ({\it i.e.} on the dimension
of the local vector space $V_j$); ii) it does not make any
assumption on the superalgebra to be considered ; iii) it is
independent of the specific functional form of the $R$-matrix
with respect to the spectral parameters: in particular, since it
does not exploit any expansion on $u-v$, it allows for
non-additive $\Rc$-matrices, which are recently object of
particular interest (see \cite{USW});  iv) it considers also the
case of `staggered' operators (the case $\sigma=-1$ below). Our
results also improve and generalize some arguments provided in
\cite{USW} and \cite{GOMU}, with which we
shall compare in the following.\\

Let us consider a single-site homogeneous operator
${\mathcal{X}}_j$; as discussed in section II it is possible to
write ${\mathcal{X}}_j=X^{\alpha}_{\beta}\,
{{\mathcal{E}}_{j}}^{\beta}_{\alpha}$, where
(see(\ref{local-local})) the matrix $X$ is actually its {\it
local} representing matrix~(\ref{local}); we obviously have that
$p(X)=p({\mathcal{X}}_j)$. We shall show that
\begin{eqnarray}
\left[ \, \check{R}(u,v), \, X \otimes^s {\Bbb I} \right. \, + &
\, & \left. \sigma \, {\Bbb I}\, {\otimes}^s X \, \right] = 0
\hspace{1cm} \sigma= \pm 1
\label{local-symm} \\
\hspace{0.1cm} \Downarrow & \, &  \nonumber \\
\hspace{0.1cm} [ \, {\mathcal{J}}_n , \sum_{i=1}^{N} \sigma^j &
\, & {\mathcal{X}}_j \, ] = 0 \label{glob-symm}
\end{eqnarray}
To prove this we shall proceed in four steps:
%%%%%%%%%%%%%%%%%%%%%%%%%%%
\\{\it \noindent Step 1}) The condition (\ref{local-symm}) implies
the following relation for the $\mathcal{L}$-operator entries:
\begin{equation}
{{\mathcal{L}}_j}^{\alpha}_{\beta'}\, X^{\beta'}_{\beta} + \sigma
{{\mathcal{L}}_j}^{\alpha}_{\beta}  \, (-1)^{p(X) p(\beta)} \,
{{\mathcal{X}}_j} \, = \, {{\mathcal{X}}_j}
\,{{\mathcal{L}}_j}^{\alpha}_{\beta} (-1)^{p(X) p(\alpha)} +
\sigma {{\mathcal{L}}_j}^{\delta'}_{\beta}\, X^{\alpha}_{\delta'}
\quad \forall j \label{1->2}
\end{equation}
where the dependence on $u$ and $v$ has been dropped to simplify
the notation.
\\{\it \noindent Proof:} It is sufficient to develop (\ref{local-symm})
in its matrix entries ({\it e.g.} the rows $(\gamma,\alpha)$ and
columns $(\beta,\delta)$) by means of  (\ref{tp^s}); then, by
using the fact that $\check{R}^{\gamma \alpha}_{\beta \delta}=
(-1)^{p(\alpha) p(\gamma)} \tilde{R}^{\alpha\gamma}_{\beta
\delta}$ and that the local $d \times d$ matrix $X$ has a
definite parity $p(X)$, one gets that eqn.(\ref{local-symm}) is
equivalent~to
\begin{equation}
\tilde{R}^{\alpha \gamma}_{\beta' \delta} \, X^{\beta'}_{\beta} +
\sigma \tilde{R}^{\alpha \gamma}_{\beta \delta'} \,
X^{\delta'}_{\delta} (-1)^{p(X)p(\beta)} = (-1)^{p(X)p(\alpha)}
X^{\gamma}_{\beta'} \, \tilde{R}^{\alpha \beta'}_{\beta \delta}
\,+ \sigma X^{\alpha}_{\delta'} \, \tilde{R}^{\delta'
\gamma}_{\beta \delta}  \label{local-symm-mia}
\end{equation}
Multiplying this equation by ${{\mathcal{E}}_j}^{\delta}_{\gamma}$
and summing up over $\gamma$ and $\delta$, one easily arrives at
(\ref{1->2}) with the help of the identities $
{{\mathcal{L}}_j}^{\alpha}_{\beta'} \, {{\mathcal{X}}_j}=
\tilde{R}^{\alpha\gamma}_{\beta\delta'} \, X^{\delta'}_{\delta}
\, {{\mathcal{E}}_j}^{\delta}_{\gamma} \, $ and $
{{\mathcal{X}}_j} \, {{\mathcal{L}}_j}^{\alpha}_{\beta}=
\tilde{R}^{\alpha \beta'}_{\beta \delta} X^{\alpha'}_{\beta'} \,
{{\mathcal{E}}_j}^{\delta}_{\alpha'} \nonumber $.
%%%%%%%%%%%%%%%%%%%%%%%%%%%%
\\{\it \noindent Step 2}) Using (\ref{1->2}) one can show that
a similar relation holds for the monodromy matrix,~{\it i.e.}
\begin{equation}
{{\mathcal{T}}_{N}}^{\alpha}_{\beta'} \, X^{\beta'}_{\beta} +
{{\mathcal{T}}_{N}}^{\alpha}_{\beta} \, (-1)^{p(X) p(\beta)} \,
\sum_{j=1}^{N} \sigma^j {{\mathcal{X}}_j}   = \sum_{j=1}^{N}
\sigma^j {{\mathcal{X}}_j} \cdot \,
{{\mathcal{T}}_{N}}^{\alpha}_{\beta} (-1)^{p(X) p(\alpha)} +
\sigma^N {{\mathcal{T}}_{N}}^{\delta'}_{\beta} \,
X^{\alpha}_{\delta'} \label{2->3}
\end{equation}
{\it Proof:} This can be done by induction; supposing that
(\ref{2->3}) holds for chain with $N$ sites, it can be seen that
is also holds for a chain with $N+1$ sites (for $N=1$
eqn.(\ref{2->3}) is nothing but eqn.(\ref{1->2}) itself). Indeed,
it is sufficient to multiply eqn.(\ref{2->3}) by
${{\mathcal{L}}_{N+1}}^{\alpha'}_{\alpha}$ on the left, sum up
over $\alpha$ and use eqn.(\ref{1->2}) with $j\rightarrow N+1$;
finally, since eqn.(\ref{L-op-sc}) implies that
\begin{eqnarray}
{{\mathcal{L}}_{N+1}}^{\alpha'}_{\alpha} \,
\sum_{j=1}^{N}{{\mathcal{X}}_j} =(-1)^{p(X)
(p(\alpha)+p(\alpha'))} \,  \sum_{j=1}^{N}{{\mathcal{X}}_j} \,
{{\mathcal{L}}_{N+1}}^{\alpha'}_{\alpha}
\nonumber \\
{{\mathcal{X}}_{N+1}} \, {{\mathcal{T}}_{N}}^{\delta'}_{\beta}=
(-1)^{p(X)\,(p(\delta')+p(\beta))}
{{\mathcal{T}}_{N}}^{\delta'}_{\beta} \,{{\mathcal{X}}_{N+1}}
\hspace{1.7cm} \nonumber
\end{eqnarray}
one easily obtains (\ref{2->3}) with $N\rightarrow N+1$.
%%%%%%%%%%%%%%%%%%%%%%%%%
\\{\it \noindent Step 3}) Eqn.(\ref{2->3}) implies that
\begin{eqnarray}
\mbox{if }  \sigma=+1
 \, \, \Rightarrow \,
 \, & \left[ \tau(u,v) \, , \,
\sum_{i=1}^{N} {{\mathcal{X}}_j} \right] = 0 \hspace{1cm} \,
& \, \, \forall \, u,v \label{3->4_1} \\
\mbox{if }  \sigma=-1
 \, \, \Rightarrow \,
 \, & \{ \tau(u,v) \, , \,
\sum_{i=1}^{N} (-1)^j {{\mathcal{X}}_j} \} = 0 \, & \, \, \forall
\, u,v \label{3->4_2}
\end{eqnarray}
{\it Proof:} In fact, if $p(X)=0$ one takes the {\it supertrace}
of (\ref{2->3}), whereas if $p(X)=1$ one can take the ordinary
{\it trace} of (\ref{2->3}); in both cases\footnote{in the case
$\sigma=-1$ one has to assume that the number of sites is even;
this is quite customary in systems that fulfill symmetries of the
kind $\sum_{i=1}^{N} (-1)^j {{\mathcal{X}}_j}$ (see for instance
\cite{YANG}).} one obtains equations involving the super-trace of
the monodromy matrix, namely (\ref{3->4_1})-(\ref{3->4_2}).
%%%%%%%%%%%%%%%%%%%%%%%%%%%%%%%%%%%%
\\{\it \noindent Step 4}) Both eqn.(\ref{3->4_1}) and
(\ref{3->4_2}) imply that
\begin{equation}
\left[ \tau^{-1}(u,w) \, \tau(v,w) \, , \, \sum_{i=1}^{N}
\sigma^{j} {{\mathcal{X}}_j} \right] = 0
\end{equation}
from which eqn.(\ref{glob-symm}) is easily deduced due to
eqn.(\ref{J_n}).
\\{\it \noindent Proof:} For the case $\sigma=+1$ the proof
is trivial; for the case $\sigma=-1$ one can observe that
eqn.(\ref{3->4_2}) implies that $ \{ \tau^{-1}(u,v),
\sum_{i=1}^{N} (-1)^{j} {{\mathcal{X}}_j} \} = 0 $, from which
\begin{displaymath}
[ \tau^{-1}(u,w) \, \tau(v,w) , \sum_{i=1}^{N} (-1)^{j}
{{\mathcal{X}}_j} ] = 2 \tau^{-1}(u,w) \, \{ \tau(v,w),
\sum_{i=1}^{N} (-1)^{j} {{\mathcal{X}}_j} \} = 0
\end{displaymath}
This concludes the proof. It is worth pointing out that the
relations obtained in the cases $\sigma=1$ and $\sigma=-1$ (see
eqns.(\ref{3->4_1}) and (\ref{3->4_2}) respectively) both involve
the supertrace of the monodromy matrix. It would not be possible
to obtain in general similar relations for the ordinary trace,
contrary to what eqn.(\ref{Z}) could suggest. This simply
confirms the role of the supertrace in fermionic systems;
nevertheless, a more detailed inspection of the third step shows
that in the case of {\it even operators} ${\mathcal{X}}_j$,
besides eqns.(\ref{3->4_1})-(\ref{3->4_2}), one also has
\begin{eqnarray}
\mbox{if }  \sigma=+1
 \, \, \Rightarrow \,
 \, & \left[ tr \, {\mathcal{T}}(u,v) \, , \,
\sum_{i=1}^{N} {{\mathcal{X}}_j} \right] = 0 \hspace{1cm} \,
& \, \, \forall \, u,v  \label{cons-tr-1} \\
\mbox{if }  \sigma=-1
 \, \, \Rightarrow \,
 \, & \{ tr \, {\mathcal{T}}(u,v) \, , \,
\sum_{i=1}^{N} (-1)^j {{\mathcal{X}}_j} \} = 0 \, & \, \, \forall
\, u,v  \label{cons-tr-2}
\end{eqnarray}
meaning that, as long as only even operators are dealt with, the
ordinary trace plays a perfectly analogous role as the supertrace.

Finally, it is worth stressing that eqn.(\ref{local-symm}) can be
rewritten in terms of the fermionic $\Rc$-matrix, obtaining the
equation $ \left[ {\check{\mathcal{R}}}_{j,j+1}(u,v) ,
{\mathcal{X}}_j + \sigma {\mathcal{X}}_{j+1} \right] \, =  \, 0
\label{local-symm-op} $. Such form is quite appealing because the
symmetries determined by $\Rc$ are immediately readable at least
for the first constant of motion ${\mathcal{J}}_1$,
{\it i.e.} the Hamiltonian, thanks to eqn.(\ref{Ham-from-Rc-op}).\\

We now want to point out that, for the subcase $\sigma=+1$, the
result (\ref{glob-symm}) was obtained in \cite{GOMU} starting not
from (\ref{local-symm}) but from a different kind of local
condition on the $R$-matrix, which we report here:
\begin{eqnarray}
\tilde{R}^{\alpha \gamma}_{\beta' \delta}(u,v) \,
X^{\beta'}_{\beta} + \tilde{R}^{\alpha \gamma}_{\beta
\delta'}(u,v) \, X^{\delta'}_{\delta} =
\label{local-symm-Goe-Mura} \hspace{7cm} \\
= (-1)^{p(X)(p(\alpha')+p(\beta))} X^{\alpha}_{\alpha'} \,
\tilde{R}^{\alpha' \gamma}_{\beta \delta}(u,v) \,+
(-1)^{p(X)(p(\alpha)+p(\beta))} X^{\gamma}_{\gamma'} \,
\tilde{R}^{\alpha \gamma'}_{\beta \delta}(u,v)  \nonumber
\end{eqnarray}
where as usual $\tilde{R}^{\alpha \beta}_{\gamma \delta}=
(-1)^{\alpha \gamma} \Rc^{\beta \alpha}_{\gamma \delta}$.
\\Such an equation was proposed \cite{GOMU} just in this form
(it cannot be set in a compact form for $\check{R}$) to explicitly
suggest that, since the $\mathcal{L}$-operator is constructed
with $\tilde{R}$ (see eqn.(\ref{L-op})), the local symmetry
constraints should be imposed on $\tilde{R}$, and not on
$\check{R}$. However, it is worth emphasizing that, due to the
fact that the constants of motion are derived from~${\mathcal{Z}}$
(see eqn.(\ref{Z})) and not directly from the transfer matrix
$\tau$, the Hamiltonian is directly related to $\check{R}$ and
not to $\tilde{R}$ (see eqn.(\ref{2-site-Ham-mat2})). Indeed the
constraint appearing in equation (\ref{local-symm}) is the natural
requirement suggested by $[ {\mathcal{H}}_{j,j+1} ,
{\mathcal{X}}_j + {\mathcal{X}}_{j+1} ] \, =  \, 0 $. Anyway,
unlike eqn.(\ref{local-symm-Goe-Mura}), eqn.(\ref{local-symm})
with $\sigma=+1$ can by equivalently imposed on $\Rc$ or on $R$.
Moreover, it must also be observed that not any operator
${{\mathcal{X}}_j}$ can be used to implement the
condition~(\ref{local-symm-Goe-Mura}) on the $R$-matrix. In fact,
if one evaluates eqn.(\ref{local-symm-Goe-Mura}) for
$(u,v)=(u_0,v_0)$ and takes into account the fact that for such
values we have $\tilde{R}^{\alpha \beta}_{\gamma
\delta}=(-1)^{\alpha \gamma} \delta^{\alpha}_{\delta}\,
\delta^{\gamma}_{\beta}$ (see eqn.(\ref{cond-2})), the following
relation is obtained
\begin{equation}
X^{\gamma}_{\beta} \delta^{\alpha}_{\delta} \left(1-(-1)^{p(X)
\,p(\beta)}) \right) + X^{\alpha}_{\delta} \delta^{\gamma}_{\beta}
\left(1-(-1)^{p(X) \,p(\delta)}) \right) = 0 \label{Goe-Mura-2}
\end{equation}
Eqn.(\ref{Goe-Mura-2}) is identically satisfied when $p(X)=0$; on
the contrary, when ${{\mathcal{X}}_j}$ is an odd operator, we can
observe that there must exist at least a couple of indices
${\gamma,\beta}$ with $p(\beta) \neq p(\gamma)$ for which
$X^{\gamma}_{\beta}\neq 0$; taking $\alpha=\delta$ in
eqn.(\ref{Goe-Mura-2}) we obtain that
$X^{\gamma}_{\beta}\,(1-(-1)^{\beta})=0$.  This means that
eqn.(\ref{local-symm-Goe-Mura}) only allows for those odd
matrices $X$ that have non-vanishing entries in {\it odd rows}
and {\it even columns}, which is a somehow `asymmetric'
constraint. This would not allow, for instance, all the odd
operators of the $u(2,2)$ superalbegra of the EKS-model (see
section VII for more details).
\\On the contrary, if one
applies the same argument to eqn.({\ref{local-symm}) -- or
equivalently to eqn.(\ref{local-symm-mia}) --, no constraints are
obtained on $X$, so that any kind of odd operator
${\mathcal{X}}_j$ can be used, as observed above; moreover, one
can easily see that, if $X$ fulfills the equation
(\ref{local-symm}), then $X^{\dagger}$ does as well, as it is
expected to be the case when exploiting the hermiticity of the
Hamiltonian in $[ {\mathcal{H}}_{j,j+1} , {\mathcal{X}}_j +
{\mathcal{X}}_{j+1} ] \, =  \, 0 $. Therefore the condition
(\ref{local-symm}) seems to be more
general than eqn.(\ref{local-symm-Goe-Mura}).\\

Finally, we wish to comment on the physical implication of the
above theorem. In doing that we shall anticipate some notions on
the extended Hubbard models that will be widely treated in the
following. The reader non familiar with these models can find
details in section VII or in the references henceforth given.
\\We first of all want to stress that the above theorem (\ref{local-symm})
$\Rightarrow$ (\ref{glob-symm}) generalizes the result obtained
by Umeno, Shiroishi and Wadati \cite{USW} on the ordinary Hubbard
model. Indeed it is known that at half filling the Hamiltonian
enjoys the $so(4)$ symmetry given by two orthogonal $su(2)$
sectors; the former is given by the spin, and the latter by the
operators $\eta^{-}= \sum_{i=1}^{N} c^{}_{i \downarrow} c^{}_{i
\uparrow}$, $\eta^{+}= (\eta^{-})^{\dagger}$, and
$\eta^{z}=\sum_{i=1}^{N} ( n_{i \uparrow} + n_{i \downarrow}
-1)/2$. Using the technique of the fermionic $R$-matrix, the above
authors not only realized that the $\check{\mathcal{R}}$-matrix
fulfills the $so(4)$-symmetry, but also showed that all the
constants of motion enjoy such a symmetry; for the generators
$\eta^{\pm}$ the prove was just based on realizing that $\{ str
\, {\mathcal{T}}_{N}(u,v), \eta^{\pm} \}=0$. In pass, we also
remark that, since $\eta^{\pm}$ are even operators, one could
also obtain that $\{ tr \, {\mathcal{T}}_{N}(u,v), \eta^{\pm}
\}=0$, as we observed above (see eqns(\ref{cons-tr-1}) and
(\ref{cons-tr-2})). A slight modification of the proof supplied
in \cite{USW} actually confirms that.
\\Secondly, it is also worth remarking that, if
the $\Rc$-matrix is imposed to commute with the number of
particles ({\it i.e.} with ${\mathcal{X}}_i = n_{i}$), all the
${\mathcal{J}}_n$ preserve the total number of particles; then
the eigenvectors of the Fock space determined by the
${\mathcal{J}}_n$ belong to fixed-$N$ subspaces; this is quite
important because it ensures that such eigenstates can also be
given a first quantization expression in terms of wave functions,
which was not obvious a priori since the problem is formulated in
the
non-fixed particle number language of second quantization.\\
\label{sec-6}
%%%%%%%%%%%%%%%%%%%%%%%%%%%%%%%%%%%%
%%%%%%%%%%%%%%%%%%%%%%%%%%%%%%%%%%%%
%%%%%%%%%%%%%%%%%%%%%%%%%%%%%%%%%%%%
\section{\boldmath The technique of polynomial $\Rc$-matrices
and its role with respect to symmetries} The Polynomial
$\Rc$-matrix Technique (PRT) is somehow complementary (with
respect to symmetries) to the theorem proved in sec.VI. Indeed
the latter exploits the symmetries of the $\Rc$-matrix to deduce
the symmetries of the the constants ${\mathcal{J}}_n$ (and in
particular of the Hamiltonian); the PRT is a constructive method
allowing to look for solutions of the YBE starting with a given
Hamiltonian of interest; in particular if ${\mathcal{H}}$ belongs
to a certain class of symmetry, the PRT precises sufficient
conditions under which the $\Rc$-matrix fulfills the same
symmetries as ${\mathcal{H}}$. Using the PRT one straightforwardly
obtain the result that, if the $\Rc$-matrix is a first or a
second order polynomial, any symmetry imposed on the Hamiltonian
immediately reflects onto the $\Rc$-matrix. This issue is of
great interest for fermionic models, in that all the known models
(apart from the ordinary Hubbard model, which has very peculiar
$\Rc$-matrix, in that it depends on two spectral parameters) have
$\Rc$-matrices that are first or second degree polynomials. We
shall come again to this topic in next section.
\\Here we just briefly recall the main aspects of this method,
which we shall use  in combination with the theorem of sec.~VI,
in order to get further information about the symmetries. We
consider for simplicity the case of additive $\Rc$-matrices, {\it
i.e.} matrices that depend on the spectral parameters through
their difference, and search for polynomial solutions of the~YBE
\begin{equation}
\Rc(u)={\Bbb I}+ u \Rc^{(1)} + ... + \frac{u^p}{p!} \Rc^{(p)}
\label{exp-Rc}
\end{equation}
where $\Rc^{(i)} \, i=1, \ldots, p$ are matrices that do not
depend on the spectral parameters; in the case of spin-1/2
fermions and 4-dimensional local spaces they are $16 \times 16$
matrices. \\Inserting the expansion (\ref{exp-Rc}) into the~YBE
(\ref{YBE-Rc-comp}) one obtains a hierarchy of equations for the
$\Rc^{(i)}$s'. The advantage is that such equations are {\it
algebraic} and not functional. They exhibit some interesting
features (see~\cite{DOMO1}); first of all the highest degree term
($\Rc^{(p)}$ in~(\ref{exp-Rc})) must fulfill~\cite{DOMO2} the
symmetric group equations
\begin{equation}
\Rc_{23}^{(p)} \, \Rc_{12}^{(p)} \,\Rc_{23}^{(p)} =
\Rc_{12}^{(p)}\, \Rc_{23}^{(p)}\, \Rc_{12}^{(p)} \hspace{2cm}
\left(\Rc_{}^{(p)} \right )^2  \propto  {\Bbb I}  \label{SGE}
\end{equation}
In particular this implies that, for {\it first} degree
polynomial $\Rc$-matrices, the Yang-Baxter Equation is equivalent
to the Symmetric Group equations.
\\In addition, the second degree coefficient $\Rc^{(2)}$ is
always explicitly given in terms of the first degree one
$\Rc^{(1)}$ as follows
\begin{equation}
\Rc^{(2)}=\left( \Rc^{(1)} \right)^2 \, + \, \gamma \, {\Bbb I}
\label{sec-ord}
\end{equation}
where $\gamma$ is a ${\Bbb C}$-number.  We recall that the first
derivative of $\Rc(u)$ with respect to spectral parameter must
coincide with the 2-site Hamiltonian and therefore $\Rc^{(1)}$
must be the representing matrix of the 2-site hamiltonian one is
interested in.
\\From the above discussion, it is easily seen that if $\Rc(u)$
is a first or a second degree polynomial, it always fulfills the
same symmetries of the Hamiltonian; in fact, if $\Rc(u)$ is of
first degree, it is made of the Identity and the 2-site
Hamiltonian, so that the statement is trivial; moreover, thanks
to eqn.(\ref{sec-ord}), this property also holds for second
degree polynomials.
\\Combining this observation with the theorem of section VI
one can therefore state that every time that an hamiltonian
$\mathcal{H}$ is reproduced by a first or second order polynomial
$\Rc$-matrix, any of its symmetries is shared by the whole set of
constants of motion ${\mathcal{J}}_n$. In the following we shall
apply this sinergic combination to the study of the extended
Hubbard models; in sec.VIII we deduce the symmetries of the
EKS\cite{EKS,EKS4} and the U-supersymmetric model
\cite{BEFR}$\div$\cite{RAMA2}, which are reproduced by a {\it
first} and a {\it second} degree polynomial $\Rc$ respectively.
In sec.IX we use the PRT to find integrable models that are
$so(4)$-invariant and the theorem to deduce the symmetries of
their constants of motion. The case of the AAS-model will also be
discussed. However, we wish to emphasize the generality of the
above observations, which can be applied not only to the extended
Hubbard models but also for integrable models in general.
\label{sec-7}
%%%%%%%%%%%%%%%%%%%%%%%%%%%%%%%%%%%
%%%%%%%%%%%%%%%%%%%%%%%%%%%%%%%%%%%
\section{Symmetries in the extended Hubbard Models}
The Hubbard Model is a model of interacting electrons which was
introduced to take into account the effect of correlations in
narrow band insulators; its generalizations -- the extended
Hubbard models -- are envisaged to investigate a wide number of
physical phenomena such as metal-insulator transitions\cite{HUB},
high-$T_c$ superconductivity\cite{EKS}, quantum wires\cite{LIGR},
as well as quantum computation\cite{QUA}. We shall consider here
the following class of extended Hubbard models
\begin{eqnarray}
{\cal H}^{\mbox{\tiny $EHM$}}= & -& \sum_{\langle{j},{k}\rangle,s
} [t-X (n_{{j}, -s} + n^{}_{{k},-s})+\tilde X n^{}_{{j},-s}
n^{}_{{k},-s} ] \, c^{\dagger}_{{j},s} c^{}_{{k}, s} +  U
\sum_{j} n^{}_{{j},\uparrow}
n^{}_{{j},\downarrow} + \nonumber \\
 &+&{V\over 2}\sum_{\langle{j},{k}\rangle} n^{}_{j} n^{}_{k}
+ {W\over 2} \sum_{\langle{j},{k}\rangle,s,s' }
c^{\dagger}_{{j},s} c^{\dagger}_{{k}, s'} c^{}_{{j},{s}'}
c^{}_{{k}, s} + Y \sum_{\langle{j},{k}\rangle }
c^{\dagger}_{{j},\uparrow} c^{\dagger}_{{j},\downarrow}
c^{}_{{k},\downarrow}
c^{}_{{k}, \uparrow} + \label{EHM}  \\
&+& P \sum_{\langle{j},{k}\rangle } n^{}_{{j},\uparrow} n^{}_{{j},
\downarrow} n^{}_{k} + {Q\over 2} \sum_{\langle{j},{k}\rangle }
n^{}_{{j},\uparrow} n^{}_{{j}, \downarrow} n^{}_{{k},\uparrow}
n^{}_{{k}, \downarrow} + \, \mu \sum_{{j},s} n^{}_{{j},s}
\nonumber
\end{eqnarray}
\noindent where $c^{\dagger}_{{j},s}$ and $c^{}_{{j},s} \,$ are
the usual fermionic operators (see (\ref{antic})). Since
electrons are spin $1/2$ fermions, the variable $s$ can now take
two possible values $s=\{ \uparrow , \downarrow\}$; the subscript
$j$ labels the sites of a lattice $\Lambda$, because the creation
and annihilation operators are thought of as referring to the set
of Wannier functions. Moreover $\, n_{{j},s} =
c^{\dagger}_{{j},s} c^{}_{{j},s}$ and $n^{}_{j} = n^{}_{j \,
\uparrow}+n^{}_{j\, \downarrow}$. The symbol $\langle {j} , \,
{k} \rangle$ stands for ordered couples of nearest neighbors in
$\Lambda$. In~(\ref{EHM}) the term $t$ represents the band energy
of the electrons, while the subsequent terms describe their
Coulomb interaction energy in a narrow band approximation: $U$
parametrizes the on-site diagonal interaction, $V$ the
neighboring site charge interaction, $X$ the  bond-charge
interaction, $W$ the exchange term, and $Y$ the pair-hopping
term. Moreover, additional many-body coupling terms have been
included  in agreement with \cite{dBKS}: $\tilde X$ correlates
hopping with on-site occupation number, and $P$ and $Q$ describe
three and four-electron interactions. Finally $\mu$ is the
chemical potential. The class (\ref{EHM}) depends on 10
parameters and is quite wide; in the literature some even more
general class of models have been considered (see for instance
\cite{BKZ,ALBA}) in which the coupling constants are
spin-dependent; however, it can be proved that the Hamiltonian
(\ref{EHM}) is the most general single-band Hamiltonian that
preserves the spin and the total charge, and that is isotropic
({\it i.e.} ${\mathcal{H}}_{j \, j+1} ={\mathcal{H}}_{j +1\,
j}$). It therefore constitutes a fairly general starting point.
The $su(2)$-spin generators explicitly read
\begin{equation}
{\mathcal{S}}^{+}=\sum_j c^{\dagger}_{j \, \uparrow} c^{}_{j \,
\downarrow} \hspace{1cm} {\mathcal{S}}^{-}=\sum_j c^{\dagger}_{j
\, \downarrow} c^{}_{j \, \uparrow} \hspace{1cm}
{\mathcal{S}}^{z}= \frac{1}{2}  \sum_j (n^{}_{j \, \uparrow} -
n^{}_{j \, \downarrow}) \label{spin}
\end{equation}
while the $u(1)$-charge generator is simply given by the operator
\begin{equation}
\eta^{z}= \frac{1}{2}  \sum_j (n^{}_{j \, \uparrow} + n^{}_{j \,
\downarrow}-1) \label{charge}
\end{equation}
In this section we shall investigate for this Hamiltonian several
symmetries of the kind envisaged in section VI, {\it i.e.} we
shall impose that ${\mathcal{H}}^{\mbox{\tiny $EHM$}}$ commutes
with a global generator in the form ${\mathcal{X}}=\sum_{j}
\sigma^j {\mathcal{X}}_j$, where the single-site operators
${\mathcal{X}}_j$ are the generators of a local algebra or
superalgebra; the $\sigma=\pm 1$ accounts for possible relative
signs between the local generators of neighboring sites.
\\In this way we find all the mutual relations
that the parameters $(t,X,\tilde{X},U,V,W,Y,P,Q,\mu)$ must
satisfy in order for these constraints to be fulfilled. To this
purpose the matrix representation of fermionic operators
developed in sec.II turns out to be a very useful tool, since it
allows to study the symmetries directly on the ${\Bbb C}$-number
matrices representing the Hamiltonian and the generators, so that
one is reconducted use techniques of ordinary algebra. In
particular, due to the fact that the ${\mathcal{H}}^{\mbox{\tiny
$EHM$}}$ involves nearest-neighbor interaction terms, the global
constraint $ [ \, {\mathcal{H}}^{\mbox{\tiny $EHM$}} , \, \sum_j
\sigma^j \, {\mathcal{X}}_j ] =0 $ is equivalent to the 2-site
form
\begin{equation}
[ \, H^{\mbox{\tiny $EHM$}}_{\mbox{\tiny 2 sites}}
(t,X,\tilde{X},U,V,W,Y,P,Q,\mu)\, , \, X \otimes^s {\Bbb I} +
\sigma \, {\Bbb I} \otimes^s X \, ]=0 \label{EHM-symm2}
\end{equation}
where $X$ is the $4 \times 4$ local representing matrix of the
generator ${\mathcal{X}}_j$, and $H^{\mbox{\tiny
$EHM$}}_{\mbox{\tiny 2 sites}}$ is the $16 \times 16$ matrix
representing the 2-site terms of~(\ref{EHM}), explicitly given
in~(\ref{EHM-matr}), where
\begin{eqnarray}
h^{11}_{11}&=&h^{22}_{22}=\mu+V-W  \hspace{1.3cm}
h^{12}_{12}=h^{21}_{21}=\mu+V \hspace{1.4cm}
h^{13}_{13}=h^{31}_{31}=h^{23}_{23}=h^{32}_{32}= \mu /2
\nonumber  \\
h^{12}_{21}&=&W \hspace{1.5cm} h^{34}_{43}=Y  \hspace{1.5cm}
h^{13}_{31}=h^{23}_{32}=-t \hspace{1.6cm}
h^{14}_{41}=h^{24}_{42}=t-2 X+ \tilde{X}
\nonumber \\
h^{14}_{14}&=&h^{41}_{41}=h^{24}_{24}=h^{42}_{42}=\frac{3}{2} \,
\mu +P+\frac{U}{2}+2 V-W  \hspace{1.5cm}
h^{34}_{34}=h^{43}_{43}=\mu+\frac{U}{2}  \label{EHM-entries}  \\
h^{44}_{44}&=&2 \mu+ 4 P+Q+U+4 V-2 W \hspace{1.5cm}
h^{12}_{34}=h^{12}_{43}=-h^{21}_{34}=-h^{21}_{43}=t-X  \nonumber
\end{eqnarray}
\begin{equation}
H^{\mbox{\tiny $EHM$}}_{\mbox{\tiny 2 sites}} =
{\scriptsize{\pmatrix{
%%%%%%%%%%%%%%%%%%%%%
\:h^{11}_{11} & 0 & 0 & 0 &|& 0 & 0 & 0 & 0 &|& 0 & 0 & 0 & 0 &|&
0 & 0 & 0 & 0 &\cr
%%%%%%%%%%%%%%%%%
\:0 & h^{12}_{12} & 0 & 0 &|& h^{12}_{21} & 0 & 0 & 0 &|& 0 & 0 &
0 & h^{12}_{34} &|& 0 & 0 & h^{12}_{43} & 0 &\cr
%%%%%%%%%%%%%%%%%%%
\:0 & 0 & h^{13}_{13} & 0 &|& 0 & 0 & 0 & 0 &|& h^{13}_{31} & 0 &
0 & 0 &|& 0 & 0 & 0 & 0 &\cr
%%%%%%%%%%%%%%%%%%%
\:0 & 0 & 0 & h^{14}_{14} &|& 0 & 0 & 0 & 0 &|& 0 & 0 & 0 & 0 &|&
h^{14}_{41} & 0 & 0 & 0 \cr
%%%%%%%%%%%%%%%%%%%%
- & - & - & - &|& - & - & - & - &|& - & - & - & - &|& - & - & - &
- &\cr
%%%%%%%%%%%%%%%%%
0 & h^{12}_{21} & 0 & 0 &|& h^{21}_{21} & 0 & 0 & 0 &|& 0 & 0 & 0
& h^{21}_{34} &|& 0 & 0 & h^{21}_{43} & 0 &\cr
%%%%%%%%%%%%%%%%%%%%
0 & 0 & 0 & 0 &|& 0 & h^{22}_{22} & 0 & 0 &|& 0 & 0 & 0 & 0 &|& 0
& 0 & 0 & 0 &\cr
%%%%%%%%%%%%%%%%%%%
0 & 0 & 0 & 0 &|& 0 & 0 & h^{23}_{23} & 0 &|& 0 & h^{23}_{32} & 0
& 0 &|& 0 & 0 & 0 & 0 &\cr
%%%%%%%%%%%%%%%%%%%
0 & 0 & 0 & 0 &|& 0 & 0 & 0 & h^{24}_{24} &|& 0 & 0 & 0 & 0 &|& 0
& h^{24}_{42} & 0 & 0 \cr
%%%%%%%%%%%%%%%%%%%%
- & - & - & - &|& - & - & - & - &|& - & - & - & - &|& - & - & - &
- &\cr
%%%%%%%%%%%%%%%%%
0 & 0 & h^{13}_{31} & 0 &|& 0 & 0 & 0 & 0 &|& h^{31}_{31} & 0 & 0
& 0 &|& 0 & 0 & 0 & 0 &\cr
%%%%%%%%%%%%%%%%%%%%
0 & 0 & 0 & 0 &|& 0 & 0 & h^{23}_{32} & 0 &|& 0 & h^{32}_{32} & 0
& 0 &|& 0 & 0 & 0 & 0 &\cr
%%%%%%%%%%%%%%%%%%%
0 & 0 & 0 & 0 &|& 0 & 0 & 0 & 0 &|& 0 & 0 & 0 & 0 &|& 0 & 0 & 0 &
0 &\cr
%%%%%%%%%%%%%%%%%%%
0 & h^{12}_{34} & 0 & 0 &|& h^{21}_{34} & 0 & 0 & 0 &|& 0 & 0 & 0
& h^{34}_{34} &|& 0 & 0 & h^{34}_{43} & 0 \cr
%%%%%%%%%%%%%%%%%%%%
- & - & - & - &|& - & - & - & - &|& - & - & - & - &|& - & - & - &
- &\cr
%%%%%%%%%%%%%%%%%
0 & 0 & 0 & h^{14}_{41} &|& 0 & 0 & 0 & 0 &|& 0 & 0 & 0 & 0 &|&
h^{41}_{41} & 0 & 0 & 0 &\cr
%%%%%%%%%%%%%%%%%%%%
0 & 0 & 0 & 0 &|& 0 & 0 & 0 & h^{24}_{42} &|& 0 & 0 & 0 & 0 &|& 0
& h^{42}_{42} & 0 & 0 &\cr
%%%%%%%%%%%%%%%%%%%
0 & h^{12}_{43} & 0 & 0 &|& h^{21}_{43} & 0 & 0 & 0 &|& 0 & 0 & 0
& h^{34}_{43} &|& 0 & 0 & h^{43}_{43} & 0 &\cr
%%%%%%%%%%%%%%%%%%%
0 & 0 & 0 & 0 &|& 0 & 0 & 0 & 0 &|& 0 & 0 & 0 & 0 &|& 0 & 0 & 0 &
h^{44}_{44} }}} \label{EHM-matr}
\end{equation}
\\As observed above, the hamiltonian (\ref{EHM}) contains 10 free
parameters (actually 9, up to an overall multiplicative~$t$); when
imposing a symmetry on ${\mathcal{H}}^{\mbox{\tiny $EHM$}}$ the
number of independent parameters obviously reduces; for each
(super)symmetry that we consider we shall also point out how many
of them remain free and comment about it.\\

{\noindent 1) \it `doubly-occupied sites'-symmetry:} It is a
$su(2) \oplus u(1) \oplus u(1)$ algebra which preserves, aside
the spin $su(2)$ and the charge $u(1)$, the following $u(1)$
generator
\begin{equation}
{\mathcal{K}}=\sum_j \, (n_{j \, \uparrow}-\frac{1}{2}) (n_{j \,
\downarrow}-\frac{1}{2}) \label{doubly}
\end{equation}
In terms of the parameters of (\ref{EHM}) this yields just one
constraint $X=t$, leaving therefore 8 free parameters (apart from
an overall multiplicative factor). This symmetry has been
exploited for instance in\cite{ALAR,SCHAD} to obtain the phase
diagram of the AAS-model (the subcase $\tilde{X}=V=W=Y=P=Q~=~0$).\\

{\noindent 2) \it $so(4)$-symmetries:} It is formed by two
mutually commuting $su(2)$ algebras; six generators are therefore
involved: three of them are the usual spin components
(\ref{spin}), while the remaining ones form another $su(2)$
sub-algebra and read
\begin{equation}
\eta^{+}_{(\sigma)}=\sum_j \sigma^{j} \, c^{\dagger}_{j \,
\downarrow} c^{\dagger}_{j \, \uparrow} \hspace{1cm}
\eta^{-}_{(\sigma)}=\sum_j \sigma^j \, c^{}_{j \, \uparrow}
c^{}_{j \, \downarrow} \hspace{1cm} \eta^{z}= \frac{1}{2}  \sum_j
(n^{}_{j \, \uparrow} + n^{}_{j \, \downarrow} -1) \label{eta}
\end{equation}
The case $\sigma=-1$ (which we shall denote by $so(4)_{(-1)}$)
was first considered by Yang\cite{YANG} for the ordinary Hubbard
model to investigate the symmetry at half-filled band. The case
$\sigma=~+1$ (denoted henceforth by $so(4)_{(+1)}$) has been
envisaged for many models~\cite{EKS,ALAR,SCHAD,FLR}. The
generators of both $so(4)_{(-1)}$ and $so(4)_{(+1)}$ are also
employed \cite{YANG,SCHAD} to build up states having the property
of Off-Diagonal Long Range Order, which is proved to imply
superconductivity~\cite{YANG2}. It can be proved that, in order
for (\ref{EHM-symm2}) to hold, the parameters must fulfill the
following mutual relations
\\{$so(4)_{(-1)}$} (5 free parameters)
\begin{equation}
{\tilde{X}}=2 X \hspace{2.5cm}  V=(W-Y-P)/2 \hspace{1cm} \mu=Y
-\frac{U}{2} \hspace{1cm} Q=-2 P \label{par-so41}
\end{equation}
\\{$so(4)_{(+1)}$} (4 free parameters)
\begin{equation}
X=t \hspace{1cm} \tilde{X}=0 \hspace{1cm} V=(W+Y-P)/2
\hspace{1cm} \mu=-Y -\frac{U}{2} \hspace{1cm} Q=-2 P
\label{par-so42}
\end{equation}
It must be emphasized that the two fermionic realizations
$so(4)_{(-1)}$ and $so(4)_{(+1)}$ are physically deeply
different: for instance $so(4)_{(+1)}$ preserves the number of
doubly occupied sites (because $X=t$), while the $so(4)_{(-1)}$
does not need to. Also, we observe that the number of free
parameters that the two realizations allow when (\ref{EHM-symm2})
is different (respectively 5 and 4); this is because, though it
can be found a transformation mapping $so(4)_{(+1)}$ into
$so(4)_{(-1)}$, such transformation does not map
the Hamiltonian (\ref{EHM}) into itself.\\

{\noindent 3) \it  $gl(2,1)$-supersymmetries}: they are formed by
8 generators; the even sector is a $su(2) \, \oplus \, u(1)$
subalgebra, made of the 3 generators (\ref{spin}) of the spin and
that of the charge (\ref{charge}). The odd sector consists of the
following four generators
\begin{eqnarray}
{\mathcal{Q}}_{\uparrow; (\sigma)}=\sum_j \sigma^j \, [\alpha \,
(1-n_{j, \downarrow}) c^{}_{j,\uparrow} + \beta \, n_{j \,
\downarrow} c^{}_{j \, \uparrow}] \hspace{2cm}
{\mathcal{Q}}^{\dagger}_{\uparrow; (\sigma)}=
({\mathcal{Q}}_{\uparrow; (\sigma)})^{\dagger} \nonumber \\
{\mathcal{Q}}_{\downarrow; (\sigma)}=\sum_j \sigma^j \, [\alpha \,
(1-n_{j, \uparrow}) c^{}_{j,\downarrow} + \beta \, n_{j \,
\uparrow} c^{}_{j \, \downarrow}] \hspace{2cm}
{\mathcal{Q}}^{\dagger}_{\downarrow; (\sigma)}=
({\mathcal{Q}}_{\downarrow; (\sigma)})^{\dagger}
\label{ferm-gen-gl21}
\end{eqnarray}
In this case the coupling constants appearing in (\ref{EHM})
become functions of the parameters $\alpha$ and $\beta$ that
determine such a linear combination. We shall distinguish three
cases:
\\{\noindent 3a)} $\alpha \neq0$ {\it and} $\beta = 0 \, $: \,
In this case we have 3 free parameters $\tilde{X},U$ and $Q$,
whereas the other ones must assume the values
\begin{displaymath}
X=t \, \hspace{1.5cm} W=V=-\sigma t \hspace{1.2cm} Y=-\sigma
(t-\tilde{X}) \hspace{1.2cm} \mu=2 \sigma t \hspace{1.2cm} P=0
\end{displaymath}
\\{\noindent 3b)} $\alpha=0$ {\it and} $\beta \neq 0 \, $:
Also in this case the free parameters are $\tilde{X},U$ and $Q$,
the remaining ones being
\begin{displaymath}
X=t \hspace{0.6cm}  W=- \sigma (t-\tilde{X}) \hspace{0.6cm}
V=-\sigma( t-\tilde{X}) +Q \hspace{0.6cm} Y=-\sigma t
\hspace{0.6cm} \mu= - U \hspace{0.8cm}  P=-Q
\end{displaymath}
These two cases are related to atypical representations of the
superalgebra $gl(2,1)$ (see \cite{MAA,RAMA}). The most
interesting situation is therefore the following
\\{\noindent  3c)} $\alpha \neq 0$ and $\beta \neq 0 \, $:
By denoting $b=\beta/\alpha$ we obtain the relations
\begin{eqnarray}
X&=&t-b (t+ \sigma \frac{U}{2}) \hspace{1.5cm} \tilde{X}=(1-b)^2
(t+ \sigma \frac{U}{2}) \hspace{1.5cm} Y=\frac{U}{2} \label{par-gl21} \\
W&=&V=-\sigma [t-b^2 (t+ \sigma \frac{U}{2})] \hspace{1.5cm} \mu=
2 \sigma  t \hspace{1.5cm} P=Q=0 \nonumber
\end{eqnarray}
where $b$ is a non-vanishing real number. These relations yield
two free parameters as long as $1+ \sigma U/2 \neq 0$; in
particular the subcase characterized by $W=V=0$ and $1+ \sigma
U/2 \neq 0$ is known in the literature as the $U$~-supersymmetric
model, and is a 1-parameter model. On the contrary, when
$1+\sigma U/2 =0$ no free parameters are left and the only
possibility allowed is the so-called EKS-model
\begin{eqnarray}
X&=&t \hspace{0.8cm} \tilde{X}=0 \hspace{0.8cm} U=-2 \sigma t
\hspace{0.8cm} V=- \sigma t \hspace{0.8cm} W=-\sigma t \nonumber \\
Y&=& - \sigma t \hspace{1.5cm} P=0 \hspace{1.5cm} Q=0
\hspace{1.5cm} \mu= 2 \sigma t \label{Pg-Pne}
\end{eqnarray}
Notice that in this case the values of the parameters are
independent of $b$; this means that the EKS-models commute with
(\ref{ferm-gen-gl21}) for any $b \neq 0$ (see the case of
$u(2,2)$ discussed below).
\\We also want to remark the relationships between the EKS-models
and the U-supersymmetric ones; we remind that the
U-supersymmetric is the subcase $W=V=0$, while the EKS-model has
$W=-\sigma t$. First of all it is worth rewriting
eqns.(\ref{par-gl21}) only in terms of the parameters of the
Hamiltonian ({\it i.e.} eliminating $b$). To this purpose the
couple $X$-$W$ turns out to be suitable. Since for $W \neq \sigma
t$ we have two free parameters, whereas for $W= -\sigma t$ we
just have the EKS-models, the allowed values in the $X$-$W$~plane
are those that belong to the two half-planes $W/t \lessgtr
-\sigma$, with the addition of the single point $(W/t=-\sigma;
X/t=1)$ representing the EKS model. In the former case
eqns.(\ref{par-gl21}) are rewritten as (we divide by $t$ to
eliminate trivial overall factors)
\begin{eqnarray}
U / t &=& 2 \sigma \left( -1+\frac{(X/t-1)^2}{1+\sigma W/t}
\right) \hspace{1cm} \tilde{X}/t=\frac{(X/t+\sigma
W/t)^2}{1+\sigma W/t}
\hspace{1cm} P=Q=0  \nonumber \\ & & \label{par-gl21-2} \\
Y/t&=&\sigma \left( -1+\frac{(X/t-1)^2}{1+\sigma W/t} \right)
\hspace{1.5cm} \mu/t=2 \sigma \hspace{2cm} \mbox{for: } \, W/t \,
\neq \, -\sigma \nonumber
\end{eqnarray}
In the $X$-$W$ plane the $U$~-supersymmetric model is represented
by the axis $W=0$ (see fig.\ref{sym_fig1}); the relations for the
other parameters are the following\footnote{The first relation of
(\ref{par-U-ss}) shows that, in spite of the name, the
U-supersymmetric model is more suitably parametrized by $X$ than
by $U$. Indeed $U$ is a continuous function of $X$ (a parabola),
while $X$ is not even a single-valued function of $U$: this
drawback both forces to consider only a sub-part of the parabola,
and may also lead to non appropriate choices of the
parametrization (like in \cite{BGLZ}) that wrongly let think of
discontinuities at $U=0$ which actually do not exist.}
\begin{eqnarray}
U / t &=& 2 \sigma \left( (X/t)^2-2 X/t \right)
\hspace{1cm} \tilde{X}/t=(X/t)^2 \hspace{1cm} W=V=P=Q=0 \nonumber \\
Y/t&=&\sigma \left( (X/t)^2-2 X/t \right) \hspace{1cm} \mu/t=2
\sigma \label{par-U-ss}
\end{eqnarray}
\\This pictorial characterization of the U-supersymmetric and EKS
model also provides  an interesting geometrical construction; in
fact it can be shown \cite{MON} that any model belonging to the
2-parameter $gl(2,1)$-class (\ref{par-gl21}) can be written as a
linear combination of the EKS model and an appropriate
U-supersymmetric model ${\mathcal{H}}_{U_{ss}}$. In the $X$-$W$
variables this property can be expressed as follows:
\begin{equation}
{\mathcal{H}}_{gl(2,1)} (X,W)= [t+\sigma W] \,
{\mathcal{H}}_{U_{ss}} \, - \sigma W \, {\mathcal{H}}_{EKS}
\label{gl21-lin-comb}
\end{equation}
The interesting fact is that the appropriate model
${\mathcal{H}}_{U_{ss}}$ can be determined in a geometrical way;
given any $gl(2,1)$-invariant model (characterized by a certain
point $P$ of coordinates $(X,W)$) it is sufficient to trace the
line from $P$ to the point $E=(X/t=1;W/t=-\sigma)$ (which is the
EKS-model), and determine the  point $A$ at which the above line
intersects the axis $W=0$ (see fig.\ref{sym_fig1}). This point
precisely represents the ${\mathcal{H}}_{U_{ss}}$-model we need
in eqn.(\ref{gl21-lin-comb}). Its coupling constants can be
determined from its ascissa $X_A$ by means of the relations
(\ref{par-U-ss}).
\\Eqn.(\ref{gl21-lin-comb}) is quite important in that it explicitly
shows that the study of $gl(2,1)$-invariant models with $W
\approx -\sigma t$ actually consists in treating the
U-supersymmetric model ${\mathcal{H}}_{U_{ss}}$ as a perturbation
of the EKS-model; on the contrary, in the limit $W \approx 0$, it
is the EKS-model that acts
as a perturbation on the known model ${\mathcal{H}}_{U_{ss}}$.\\

{\noindent 4) \it $so(5)$-symmetries}: it is an algebra made of
10 generators; we shall consider the realization in terms of
fermionic operators introduced in \cite{FLR}
\begin{eqnarray}
{\mathcal{A}}^{2 3}&=&2 {\mathcal{S}}^{+} \hspace{1.9cm}
{\mathcal{A}}^{3 2}=2 {\mathcal{S}}^{-} \hspace{1.9cm}
{\mathcal{A}}^{2 2}=2 {\mathcal{S}}^{z} \hspace{4cm} \nonumber
\\
{\mathcal{A}}^{1 4}&=&2 \eta^{+}_{(1)} \hspace{1.8cm}
{\mathcal{A}}^{4 1}=2 \eta^{-}_{(1)} \hspace{1.9cm}
{\mathcal{A}}^{1 1}=2 \eta^{z} \hspace{4cm} \label{ferm-gen-so5}
\\
{\mathcal{A}}^{1 2}_{(\sigma)}&=&\sum_j \sigma^j c^{\dagger}_{j
\, \downarrow} \hspace{1cm} {\mathcal{A}}^{2 1}_{(\sigma)}=\sum_j
\sigma^j c^{}_{j \, \downarrow} \hspace{1cm} {\mathcal{A}}^{1
3}_{(\sigma)}=- \sum_j \sigma^j c^{\dagger}_{j \, \uparrow}
\hspace{1cm} {\mathcal{A}}^{3 1}_{(\sigma)}=-\sum_j \sigma^j
c^{}_{j \, \uparrow} \nonumber
\end{eqnarray}
This is a subcase of the algebra $so(4)_{(+1)}$ defined above;
the obtained relations among the parameters only allow the
EKS-models~(\ref{Pg-Pne}). The latter are therefore the only
models of the extended Hubbard class~(\ref{EHM}) that enjoy the
$so(5)$-symmetry; actually, there are other 1-$d$ single-band
models commuting with (\ref{ferm-gen-so5}) (see \cite{FLR}), but
they do not belong to the class~(\ref{EHM}) since they are not
isotropic. Even if such models exhibit a superconducting
behaviour, it should be observed that a more realistic
description of the superconducting transition is reached when
assuming that the Hamiltonian is isotropic and the wave function
breaks the lattice symmetries
(otherwise there is no break-up in fact).\\

{\noindent 5) \it $u(2,2)$-supersymmetry}: It is made of 16
generators; the even sector contains the 3 generators of the spin
and the 3 ones of the $so(4)_{(+1)}$ algebra, as well as the
identity and the doubly-occupied sites generator (\ref{doubly}).
The odd sector contains the following eight generators
\begin{eqnarray}
\hat{\mathcal{Q}}_{\uparrow; (\sigma)} &=& \sum_{j} \sigma^j \,
(1-n^{}_{j \, \downarrow}) \, c^{}_{j \, \uparrow} \hspace{1cm}
\hat{\mathcal{Q}}_{\downarrow; (\sigma)} = \sum_{j} \sigma^j \,
(1-n^{}_{j \, \uparrow}) \, c^{}_{j \, \downarrow} \hspace{1cm}
\hat{\mathcal{Q}}^{\dagger}_{\uparrow; (\sigma)} \hspace{1cm}
\hat{\mathcal{Q}}^{\dagger}_{\downarrow; (\sigma)} \nonumber \\
\tilde{{\mathcal{Q}}}_{\uparrow; (\sigma)} &=& \sum_{j} \sigma^j
\, n^{}_{j \, \downarrow} \, c^{}_{j \, \uparrow} \hspace{2cm}
\tilde{{\mathcal{Q}}}_{\uparrow; (\sigma)} = \sum_{j} \sigma^j \,
n^{}_{j \, \downarrow} \, c^{}_{j \, \uparrow} \hspace{2cm}
\tilde{{\mathcal{Q}}}^{\dagger}_{\uparrow; (\sigma)} \hspace{1cm}
\tilde{{\mathcal{Q}}}^{\dagger}_{\downarrow; (\sigma)}
\label{ferm-gen-u22}
\end{eqnarray}
The commutations with its generators only allow the solution
(\ref{Pg-Pne}) of the EKS models. If one considered the algebra
$so(4)_{(-1)}$ instead of $so(4)_{(+1)}$
in the even sector, no extended Hubbard model (\ref{EHM}) would be found.\\

We wish to make some remarks about the EKS-models, defined by
eqn.(\ref{Pg-Pne}). The original EKS-model introduced
in~\cite{EKS} corresponds to the case $\sigma=+1$ of
eqn.(\ref{Pg-Pne}); however, we have unified $\sigma=\pm 1$ in
the definition since they often share the same kind of symmetry.
Here we precise the conditions for this to happen. First of all,
it is easily shown that the 2-site term of the EKS-models are
respectively a graded permutator\cite{EKS} (the case $\sigma=+1$,
denoted ${\mathcal{P}}^g$), and an oppositely-graded permutator
(the case $\sigma=-1$, denoted $\bar{\mathcal{P}}^g$): their $16
\times 16$ matrix representations are $ (P^g)^{\alpha
\beta}_{\gamma \delta} =(-1)^{p(\alpha) p(\beta)} \,
\delta^{\alpha}_{\delta} \delta^{\beta}_{\gamma} $ and $
(\bar{P}^g)^{\alpha \beta}_{\gamma \delta}=(-1)^{(p(\alpha)+1)
(p(\beta)+1)} \,\delta^{\alpha}_{\delta} \delta^{\beta}_{\gamma}
$ respectively. It can also be proved that, for any homogeneous
operator ${\mathcal{X}}_j$,
\begin{equation}
{\mathcal{P}}^g_{jk} \, {\mathcal{X}}_j \, {\mathcal{P}}^g_{jk} =
{\mathcal{X}}_k \hspace{2cm} \bar{\mathcal{P}}^g_{jk}\,
{\mathcal{X}}_j \, \bar{\mathcal{P}}^g_{jk} =
(-1)^{p({\mathcal{X}}_j)} {\mathcal{X}}_k \quad . \label{prop-op}
\end{equation}
Using (\ref{prop-op}) it is easy to realize that, if $X$ is the
local representing matrix of ${\mathcal{X}}_j$
\begin{equation}
[ P^g \, , \, X \otimes^s {\Bbb I} + {\Bbb I} \otimes^s X ] =0
\hspace{1cm} \forall \, X \, \mbox{ even and odd} \label{prop-Pg}
\end{equation}
whereas
\begin{eqnarray}
\hspace{0.1cm} [ \bar{P^g} \, , \, X \otimes^s {\Bbb I} + {\Bbb
I} \otimes^s X ] &=& 0 \hspace{1cm} \forall \, X \,
\mbox{even} \label{prop1-Pne} \\
\hspace{0.1cm} [ \bar{P^g} \, , \, X \otimes^s {\Bbb I} - {\Bbb
I} \otimes^s X ] &=& 0 \hspace{1cm} \forall \, X \, \mbox{odd}
\label{prop2-Pne}
\end{eqnarray}
These relations imply that, when investigating the commutation of
the Hamiltonian with generators of the form $\sum_i \sigma^j
{\mathcal{X}}_j$, {\it both} the EKS-models are found to fulfill
the commutations as long as additional $\sigma^j$ appears in {\it
odd} generators (see for instance the cases of $gl(2,1)$, $so(5)$
and $u(2,2)$ symmetries); in particular ${\mathcal{P}}^g$ is
obtained if $\sigma=+1$, while $\bar{\mathcal{P}}^g$ when
$\sigma=-1$. Indeed odd generators with $\sigma=+1$ can be mapped
into those with $\sigma=-1$ through the transformation $c_j
\rightarrow (-1)^j c_j$ (which leaves unaltered any even
generator).
\\On the contrary, when the sign $\sigma^j$ appears in {\it even}
generators, we always obtain ${\mathcal{P}}^g$ for the subcase
$\sigma=+1$, but not $\bar{\mathcal{P}}^g$ in the subcase
$\sigma=-1$ (see for instance the cases of the
$so(4)$~symmetries). Indeed the case of symmetries with respect
to even generators with $\sigma=-1$ is quite peculiar; in
particular the study of the generators $\eta^{\pm}_{(\sigma)}$
(\ref{eta}) with $\sigma=-1$ is physically intriguing for reasons
of energetic stability of the ground state \cite{MOCA}; we shall
investigate more deeply this aspect in
section VIII.\\

Here we want to briefly recall which are, among the several
classes of symmetry found above, the subcases that (up to now)
have been proved to be integrable. As discussed in the previous
sections, the integrability of a nearest-neighbor Hamiltonian
$\mathcal{H}$ is proved whenever it is possible to find a matrix
$\Rc(u,v)$ which solves the YBE~(\ref{YBE-Rc-comp}) and such that
its first derivative with respect to the spectral parameter
equals the representation of the local hamiltonian
${\mathcal{H}}_{i \, i+1}$ (see eqn.(\ref{2-site-Ham-mat2})).
\\Focussing on the symmetries and the supersymmetries considered
above, we remind that as to the `doubly occupied sites'-symmetry,
the $\Rc$-matrix has been found for 96 models (subcases of which
are known models). Within the 2-parameter class of $gl(2,1)$
invariant models the $\Rc$-matrix has been found for the subcases
of the EKS-models \cite{EKS4} and the 1-parameter
U-supersymmetric models~\cite{MAA,PFFR,RAMA}. For the
$so(5)$-symmetry and the $u(2,2)$-supersymmetry the two EKS
models are the only extended Hubbard models~(\ref{EHM}) that
fulfill them, as observed before. So, their $\Rc$-matrices are
the ones cited just above. As to the $so(4)$-symmetry, the
$\Rc$-matrix has been found for the half filled Hubbard
model\cite{USW}.

As anticipated in sec.VIII, there is an intriguing feature shared
by the $\Rc$-matrices of all the integrable models cited just
above (apart from the Hubbard model, which has a non-additive
$\Rc$-matrix): they are all polynomials in the difference of the
spectral parameters; more precisely, the EKS-models have {\it
first} degree polynomial $\Rc$-matrices, while the
U-supersymmetric have a {\it second} degree one. Even if they are
not usually written in the literature as polynomials (because the
form in which they are given depends on the method used to find
them as solutions of the Yang-Baxter equation), it is easy to
realize that such $\Rc$-matrices can be cast into polynomials
through a mere multiplication by a scalar function $\phi$ or
redefinition of the spectral parameters (see\cite{DOMO1})). This
fact means that all the known models (apart from ordinary Hubbard)
can be found by means of the Polynomial $\Rc$-matrix Technique
discussed in sec.VIII.
\\In particular we therefore have that all the constants of motion of
the U-supersymmetric model enjoy the $gl(2,1)$-supersymmetry. As
to the EKS models (\ref{Pg-Pne}) the conclusion is still more
general; for the case $\sigma=+1$ we have the graded permutator,
and thus, using eqn.(\ref{prop-Pg}) we deduce that the constants
of motions ${\mathcal{J}}_n$ of this models fulfill {\it any}
global symmetry $\sum_j {\mathcal{X}}_j$; for the case
$\sigma=-1$ we have an oppositely graded permutator and therefore
(due to eqns.(\ref{prop1-Pne})-(\ref{prop2-Pne})) we have that
the constants of motion ${\mathcal{J}}_n$ commute with $\sum_j
{\mathcal{X}}_j$ for any ${\mathcal{X}}$ even, and with $\sum_j
(-1)^{j} {\mathcal{X}}_j$
for any ${\mathcal{X}}$ odd.\\
Furtherly, first degree polynomial $\Rc$-matrices for 96 models
have been found in \cite{DOMO2}; all these Hamiltonians are shown
to fulfill the 'doubly occupied symmetry' $su(2) \oplus u(1)
\oplus u(1)$; now we can deduce that all the constants of motion
of these models fulfill such symmetry as well.
\\Among the symmetries envisaged in this section,
the case left to be considered on the point of view of
integrability is the one of $so(4)$-invariant models. This class
is very large (contains a number of free parameter), and in spite
of the fact that it is quite interesting on a physical point of
view, it has been investigated only partially up to now. We shall
therefore devote the whole section IX to it. The PRT will be
applied together with the theorem of section VI to find
$so(4)$-invariant models and to deduce the symmetries of their
constants ${\mathcal{J}}_n$. \label{sec-8}
%%%%%%%%%%%%%%%%%%%%%%%%%%%%%%%%%%%%
%%%%%%%%%%%%%%%%%%%%%%%%%%%%%%%%%%%%
%%%%%%%%%%%%%%%%%%%%%%%%%%%%%%%%%%%%
\section{\boldmath Integrable models with
$so(4)$-symmetries} The class of $so(4)$-invariant extended
Hubbard models (see eqns.(\ref{par-so41}) and (\ref{par-so42}))
is relevant in physics because, by exploiting its properties of
symmetry, some interesting features on the eigenstates of these
models can be easily deduced, such as the form of the correlation
functions, which exhibit a Long Range Order. It is then important
to have as many exact results as possible for such class. We
recall that it is a rather wide class, since $so(4)_{(-1)}$
allows 5 free parameters and $so(4)_{(+1)}$ 4 free parameters.
\\Among the known $so(4)$-invariant models, the most famous
is perhaps the ordinary Hubbard model at half filling\cite{YANG}
(which corresponds to the subcase $X=W=Y=P=0$ of $so(4)_{(-1)}$,
see eqs.(\ref{par-so41})). The $\Rc$-matrix for this model was
found in \cite{USW}; also it has been proved that its constants
of motion fulfill such symmetry as well. Later, the EKS-models
and the AAS model were also found, for which the phase diagrams
were derived; they enjoy the $so(4)_{(+1)}$-symmetry. Moreover,
in the model recently studied by Alcaraz and Bariev within the
Coordinate Bethe Ansatz, an $so(4)_{(-1)}$-invariant subcase
($\eta=0$ and $\epsilon=+1$ in \cite{ALBA}) can be identified; it
corresponds to the 1-parameter subclass $X=(1-\sin \theta) t$ ;
$Y=-W=t \cos \theta$ and $P=0$ of eqs.(\ref{par-so41}).\\

Motivated by the physical interest in the $so(4)$-symmetries, in
this section we shall find further models enjoying such symmetry.
More precisely, we shall make use of the polynomial $\Rc$-matrix
technique to find all the integrable $so(4)$-invariant extended
Hubbard models that are derived from a first degree
polynomial~$\Rc(u)$ in the spectral parameter $u$. We consider
here both $so(4)_{(+1)}$ and $so(4)_{(-1)}$. The number of such
models turns out to be 32, 16 with symmetry $so(4)_{(-1)}$ and 16
with $so(4)_{(+1)}$; the EKS and AAS models are shown to be among
them. In providing the $\Rc$-matrix for all of them, we shall
guarantee their integrability; moreover, by exploiting the result
of the theorem proved in section VI, we will be able to show that
not only the Hamiltonian but the whole set of constants of motion
of these models are $so(4)$-invariant. All these models have the
natural properties of preserving the total magnetization and
charge and of being isotropic, since they belong to the
class~(\ref{EHM}) by construction.
\\In order to show how to find them, we have to use as $\Rc^{(1)}$
the matrix (\ref{EHM-matr}) of the extended Hubbard models, with
the additional prescriptions (\ref{par-so41}) and (\ref{par-so42})
of the $so(4)$-symmetries, and impose that $\Rc^{(1)}$ fulfills
eqns.(\ref{SGE}). We give in the following the values of their
coupling constants, as well as the $\Rc$-matrices that yield
them. In order to simplify the notation we divide the 32 models
into four subgroups according to vanishing or non-vanishing {\it
pair hopping} and {\it exchange} amplitudes; they are denoted
${\mathcal{H}}^{(a)}$, ${\mathcal{H}}^{(b)}$,
${\mathcal{H}}^{(c)}$ and ${\mathcal{H}}^{(d)}$ respectively. The
result are presented in a compact form: the case $\sigma=+1$
gives the models that are invariant under $so(4)_{(+1)}$ whereas
$\sigma=-1$ corresponds to the $so(4)_{(-1)}$-symmetric cases.
\\{\noindent \it $1^{st}$ subgroup: ${\mathcal{H}}^{(a)}=$
8 models with $Y \neq 0$ and $W \neq 0$}
%%%%%%%%%%%%%%%%%%%%%%%%%%%%
\begin{eqnarray}
X&=&t \hspace{1cm} \tilde{X}=(1-\sigma) t \hspace{1cm}
U=2 s_1 t \hspace{1cm} V=s_1 t \hspace{1cm} W=-s_2 \nonumber \\
Y&=&\sigma  s_1 \, t  \hspace{1cm} P=-(s_1+s_2) t \hspace{1cm}
Q=2 (s_1+s_2) t  \hspace{1cm} \mu=-2 s_1 t \label{group1}
\end{eqnarray}
%%%%%%%%%%%%%%%%%%%%%%%%%%%%
\\{\noindent \it $2^{nd}$ subgroup: ${\mathcal{H}}^{(b)}=$
8 models with $Y \neq 0$ and $W=0$}
\begin{eqnarray}
X&=&t \hspace{1cm} \tilde{X}=(1-\sigma) t \hspace{1cm}
U=2 s_1 t \hspace{1cm} V=(s_1+s_2) t \hspace{1cm} W=0 \nonumber \\
Y&=&\sigma  s_1 \, t  \hspace{1cm} P=-(s_1+2 s_2) t \hspace{1cm}
Q=2 (s_1+2 s_2) t  \hspace{1cm} \mu=-2 s_1 t \label{group2}
\end{eqnarray}
%%%%%%%%%%%%%%%%%%%%%%%%%%%%
\\{\noindent \it $3^{rd}$ subgroup: ${\mathcal{H}}^{(c)}=$
8 models with $Y=0$ and $W \neq 0$}
\begin{eqnarray}
X&=&t \hspace{1cm} \tilde{X}=(1-\sigma) t \hspace{1cm}
U=4 s_1 t \hspace{1cm} V=s_1 t \hspace{1cm} W=s_2 t \nonumber \\
Y&=&0 \hspace{1cm} P=(-2 s_1+s_2) t \hspace{1cm} Q=(4 s_1-2 s_2) t
\hspace{1cm} \mu=-2 s_1 t  \label{group3}
\end{eqnarray}
%%%%%%%%%%%%%%%%%%%%%%%%%%%%
\\{\noindent \it $4^{th}$ subgroup: ${\mathcal{H}}^{(d)}=$
8 models with $Y=0$ and $W=0$}
\begin{eqnarray}
X&=&t \hspace{1cm} \tilde{X}=(1-\sigma) t \hspace{1cm}
U=4 s_1 t \hspace{1cm} V=(s_1+s_2) t \hspace{1cm} W=0 \nonumber \\
Y&=&0 \hspace{1cm} P=-2(s_1+s_2) t \hspace{1cm} Q=4 (s_1+s_2) t
\hspace{1cm} \mu=-2 s_1 t  \label{group4}
\end{eqnarray}
%%%%%%%%%%%%%%%%%%%%%%%%%%%%
\\For the Hamiltonians ${\mathcal{H}}^{(a)}$ the subcases
with $s_2=-s_1$ and $\sigma=+1$ are the two EKS models, while the
subcase $s_2=-s_1=1$ and $\sigma=-1$ is the model proposed in
\cite{ALBA} with $\eta=\theta=0$ and $\epsilon=+1$. No
U-supersymmetric model appears, since one always has $\tilde{X}
\neq X^2$. The case ${\mathcal{H}}^{(d)}$ with $s_2=-s_1$ and
$\sigma=+1$ is the AAS-model \cite{ALAR,SCHAD}.
\\The models (\ref{group1})-(\ref{group2})-(\ref{group3})-(\ref{group4})
can be obtained from the following $\Rc$-matrices
\begin{equation}
\Rc(u)={\Bbb I} + u \, (H^{(i)}_{\mbox{\tiny 2-sites}}+ s_1 \, t
{\Bbb I}) \label{Rc-abcd}
\end{equation}
where $H^{(i)}_{\mbox{\tiny 2-sites}}$ with $i=a,b,c,d \,$ is the
2-site matrix (\ref{EHM-matr}) of coupling constants
(\ref{group1}) , (\ref{group2}) , (\ref{group3}) and
(\ref{group4}) respectively \footnote{As it is clear from
eq.(\ref{2-site-Ham-mat2}), the $\Rc$-matrix (\ref{Rc-abcd})
yields ${\mathcal{H}}^{(a)}$, ${\mathcal{H}}^{(b)}$,
${\mathcal{H}}^{(c)}$ and ${\mathcal{H}}^{(d)}$ up to an
additional energy shift $s_1 t \, {\Bbb I}$, which obviously
changes nothing to the integrability of such~models.}. Since the
above $\Rc$-matrices are first degree polynomials in $u$, we have
by construction that $ [ \, \Rc(u) \, , \, X_{so(4)} \otimes^s
{\Bbb I} + \sigma \, {\Bbb I} \otimes^s X_{so(4)} \, ] = 0 $
where $X_{so(4)}$ is the local representing matrix of the
single-site generator ${{\mathcal{E}}_j}^{3}_{4}= c^{\dagger}_{j
\, \downarrow} c^{\dagger}_{j \, \uparrow}$, {\it i.e.} the $4
\times 4$ matrix $E^{3}_{4}$ (see sec.II). Thus, thanks to the
theorem of section VI, we can deduce that
\begin{displaymath}
[ \, {\mathcal{J}}_n \, , \sum_j \sigma^j c^{\dagger}_{j \,
\downarrow} c^{\dagger}_{j \, \uparrow} \, ]=0 \quad .
\end{displaymath}
\\We wish now to discuss how to deduce some physical properties of
the models found above. We shall provide here a {\it general}
scheme of techniques that allow to obtain information about the
spectrum, the phase diagram of the ground state and the behaviour
of the correlation functions for the 32 models presented above.
Details on the application of the method to some specific case
are given in separate publications \cite{DOMO3,DOMO4}. At the end
of this section we provide as an example some results obtained
for two models of group 2.
\\In the first instance we shall describe the method to
derive the spectrum of the above models; such method is based on
the observation that the Hamiltonian of the above 32 models
(\ref{group1})~$\div$~(\ref{group4}) can be proved to be of the
form:
\begin{equation}
{\mathcal{H}}^{(i)} \, = \,- \, \sum_{j} \Pi_{j,j+1} \label{GP}
\end{equation}
(up to the trivial additive term $s_1 \, t \, {\Bbb I}$) where
$\Pi_{j,j+1}$ is a 2-site generalized permutator. A 2-site
generalized permutator (GP}) acts on two neighboring sites
exchanging some couples of states but leaving unchanged some
other couples; more explicitly:
\begin{equation}
\Pi_{j,j+1} \, |\alpha \rangle_j \, |\beta\rangle_{j+1} =
\theta^{o}_{\alpha \beta} \, |\beta \rangle_j |\alpha
\rangle_{j+1} \, + \, \theta^{d}_{\alpha \beta}  \, |\alpha
\rangle_j \, |\beta \rangle_{j+1} \label{GP-expl}
\end{equation}
where $\theta^{o}_{\alpha\beta}$ and $\theta^{d}_{\alpha\beta}$
are two discrete-valued (0, +1 or~-1) and `complementary'
($|\theta^{o}_{\alpha\beta}|=1-|\theta^{d}_{\alpha\beta}|$)
functions that identify the exchange / non-exchange respectively;
notice that in both cases an additional sign is possible (if
$\theta^{o},\theta^{d}=-1$); moreover, $\theta^{o}_{\alpha\beta}=
\theta^{o}_{\beta\alpha}$, due to hermiticity. These functions
$\theta^{d}_{\alpha\beta}$ and $\theta^{o}_{\alpha\beta}$
completely determine the specific kind of GP one deals with.
\\In pass, it can be proved \cite{DOMO2} that within the class
of the extended Hubbard models (\ref{EHM}) there are 96 (and only
96) models that can be cast in the form (\ref{GP}), with a
suitable GP; the 32 models (\ref{group1})~$\div$~(\ref{group4})
that we have presented here constitute the subclass of all the
$so(4)$-invariant GP, and it is also possible to see that they
cannot be mapped into other GP through any unitary transformation.
\\Remarkably, recognizing that a Hamiltonian has the structure
(\ref{GP}) allows to exploit the Sutherland's Species Technique
\cite{DOMO3,SUT}, which actually consists in regarding a GP as an
{\it ordinary} permutator between the so-called Sutherland's
species (SS), the latter being groups of the $d$ local physical
states. The groups of states are not determined by the original
Fock Space but uniquely by the structure of the Hamiltonian; the
SS therefore relate to the dynamical processes involved in the
model. In the case of the extended Hubbard models (\ref{EHM}),
the physical states per site are 4, and thus the SS are always no
more than 4. Following Sutherland's notation\cite{SUT}, each
species can be classified as either `fermionic'($F$) or
`bosonic'($B$) (odd and even in \cite{DOMO3}, respectively),
according to the sign of the non-exchange part ($\theta^d=+1$ for
$B$, while $\theta^d=-1$ for $F$). In particular, within the 32
$so(4)$-models, one can distinguish the following types of SS:
\begin{eqnarray}
\mbox{\it $1^{st}$ subgroup} & \hspace{1cm}  &
F^4 \, \, ; \, \, B^2 \, F^2 \, \, ; B^4 \nonumber \\
\mbox{\it $2^{nd}$ subgroup} & \hspace{1cm}  & F^3 \, \, ; \, \,
\, B \, F^2  \, \, ;
\, \, B^2 F \, \, ; \, \, B^3 \label{grou-spe}\\
\mbox{\it $3^{rd}$ subgroup} & \hspace{1cm}  & F^3 \, \, ; \, \,
\, B \, F^2  \, \, ;
\, \, B^2 F \, \, ; \, \, B^3 \nonumber \\
\mbox{\it $4^{th}$ subgroup} & \hspace{1cm}  & B^2 \, \, ; \, \,
B F \, \, ; \, \, F^2 \nonumber
\end{eqnarray}
where $B^{n} \, F^{m}$ characterizes a model with $n$ bosonic and
$m$ fermionic SS. Different models that are recognized to be of
the same SS-types also share the same structure of Coordinate or
Algebraic Bethe Ansatz equations; one is therefore reconducted to
known problems, as the possible different signs of off diagonal
terms ($\theta^d$) are not expected to alter the structure of
such equations. For the cases $F^{P}$ or $B F^{P}$ the Bethe
Ansatz equations have been investigated \cite{SUT} within the
framework of the Coordinate Bethe Ansatz. The cases $F^P$ and
$B^P$ have been examined in \cite{KURE} within the QISM. Some
others cases ($B F^2$ and $B^2 F^2$) have the same algebraic
structure as known models ($t-J$\cite{FOKA,EKStJ} and
$EKS$\cite{EKS4,EKS} respectively). For the $BF$ type the
spectrum, up to constant terms, is that of a spinless fermions on
a chain:
\begin{equation}
\epsilon(\{n_l \})=-2 \sum_{l=0}^{L-1} \cos (\frac{2\pi l}{L}) \,
n_l
\end{equation}
where $n_l=0,1$ are quantum numbers and $L$ the length of the
periodic chain.
\\The whole spectrum can be found in all these cases.
However, it must be realized that the actual degeneracy of each
eigenvalue depends on the way the Sutherland's species are
realized in terms of physical species in each specific model.
This in turn enters the calculation of the partition function,
determining different physical features (for the discussion in
case of $BF$ models see \cite{DOMO4}).
\\Within the spectrum, the ground state is particularly worth of interest,
because these Hamiltonians are expected to well describe
materials that exhibit peculiar physics at low temperatures. For
the $F^P$ case, it can be shown \cite{SUT} that the miminum of
the energy, whose eigenvalue reads
\begin{equation}
\epsilon_0=1-\frac{2}{P} \int_{0}^{1}   dx \,
\frac{x^{\frac{1}{P}-1}}{1-x} \quad \quad , \label{eps-FP}
\end{equation}
is reached at equal densities of all fermionic species.
Interestingly such result, when applied to the $F^3$ and $F^4$
models in (\ref{grou-spe}), implies that the ground state in these
cases always contains doubly occupied sites, which are expected
to model short coherence length pairs.
\\Also, again as far as the ground state is concerned, it
has been shown \cite{DOMO3} that Sutherland's theorem (originally
formulated\cite{SUT} for ordinary permutators) can be extended to
the generalized permutators, and it is thus possible to assert
that, in the thermodynamic limit, the ground state energy (per
site) $\epsilon_0$ of a $B^n F^m$ problem is equal to that of a
$B F^m$ problem.  The latter types become therefore the only
relevant ones as to the issue of
determining $\epsilon_0$.\\

We also want to stress another important general feature of the
32 models (\ref{group1})$\div$({\ref{group4}); their Hamiltonians
${\mathcal{H}}^{(i)}$, aside the $so(4)$-symmetry, also preserve
the number of doubly occupied sites (see section VII), since they
all have $X=t$. Hence, by adding to each ${\mathcal{H}}^{(i)}$
($i=a,b,c,d$) further terms of on-site Coulomb repulsion and
chemical potential with arbitrary coupling constants,
\begin{eqnarray}
{\mathcal{H}}^{\prime \, (i)} = {\mathcal{H}}^{(i)} + U \sum_j
n^{}_{j\, \uparrow} n^{}_{j\, \downarrow} \, + \, \mu \sum_j
(n^{}_{j\, \uparrow}+ n^{}_{j\, \downarrow}) \quad , \quad
i=a,b,c,d  \quad ,\label{H_i_prime}
\end{eqnarray}
the obtained models ${\mathcal{H}}^{\prime \, (i)}$ still preserve
the number of spin-up, spin-down electrons separately, as well as
the number of doubly occupied sites. Such property helps to
diagonalize the Hamiltonian ${\mathcal{H}}^{\prime \, (i)}$ within
each subspace of given eigenvalues of ${\mathcal S}^z$, $\eta^z$
and ${\mathcal K}$; this is interesting from a physical point of
view, because one can investigate how the features of the model
change when tuning either the parameter $U$ or the filling $n$
(which can be expressed in terms of $\mu$); the former is somehow
an intrinsic energy unit for these models, and is actually the
most meaningful parameter for the systems where strong electronic
correlations are involved, as pointed out by Hubbard himself in
his original paper\cite{HUB}; the latter is also relevant in that
it can model the degree of hole doping in the material. The two
parameters $U$ and $n$ are expected to drive the transitions
between physically different phases (Metallic, Insulator,
Superconducting).
\\The $so(4)$-symmetry of the part ${\mathcal{H}}^{(i)}$ of
eq.(\ref{H_i_prime}) implies that \footnote{To simplify the
notation, we omitted any subscript $\sigma$ in denoting
${\mathcal{H}}^{(i)}$ and ${\mathcal{H}}^{\prime \, (i)}$; we
precise that, in eq.(\ref{weak}), one has to take
$\eta^{\pm}_{(+1)}$ (resp. $\eta^{\pm}_{(-1)}$) when
${\mathcal{H}}^{(i)} \,$ in ${\mathcal{H}}^{\prime \, (i)}$
contains $\sigma=+1$ ($\sigma=-1$); see
eqns.(\ref{group1})$\div$(\ref{group4}).}
\begin{equation}
[ \, {\mathcal{H}}^{\prime \, (i)} \, , \eta^{\pm}_{(\sigma)} \,
]= (U-\frac{\mu}{2}) \, \eta^{\pm}_{(\sigma)} \quad . \label{weak}
\end{equation}
Eq.(\ref{weak}) is crucial in order to implement the so called
$\eta$-pairing mechanism for the construction of eigenstates of
${\mathcal{H}}^{\prime \, (i)}$: indeed, once a reference
eigenstate $|\mbox{\it ref } \rangle$ of ${\mathcal{H}}^{\prime
\, (i)}$ is given, further eigenstates $|\Psi_{(\sigma)}\rangle$
can be found by applying~$\eta^{+}_{(\sigma)}$ on $|\mbox{\it ref
} \rangle$, namely $|\Psi_{(\sigma)}\rangle=
(\eta^{+}_{(\sigma)})^{m} |\mbox{\it ref } \rangle$. This
property allows to reconduct the calculation of the correlation
functions for the eigenstates $|\Psi_{(\sigma)}\rangle$ to those
of the reference states $|\mbox{\it ref } \rangle$, which are in
general easier to be computed. Within the class of the above 32
models, meaningful reference states can be found: the vacuum
$|0\rangle$ is always one (the eigenstates
$|\Psi_{(\sigma)}\rangle$ constructed on it are called {\it pure}
$\eta$-pair states); the eigenstates $|U=\infty \rangle$ of the
$U=\infty$ Hubbard model\cite{HUB-inf} are also reference states
whenever $W=V=0$; similarly, for the cases in which $W=V=\pm t$
the eigenstates $|t-J \rangle$ of the $t-J$-model
\cite{FOKA,EKStJ} are reference states. In these cases the
eigenstates $|\Psi_{(\sigma)}\rangle$ constructed on $|U=\infty
\rangle$ or on $|t-J \rangle$ are referred to as {\it mixed}
$\eta$-pair states. The correlation function for these reference
states have been investigated in
the literature (see for instance \cite{EKS,SCHAD,GANG}).\\

The techniques outlined above are quite general and apply to all
the 32 $so(4)$-models (\ref{group1})$\div$(\ref{group4}). As
already observed, all the models ${\mathcal H}^{(i)}$ of a given
$B^n F^m$-type share the same spectrum equation structure.
However, this does not necessarily imply that the related
${\mathcal H}^{\prime \, (i)}$ have all the same physical
features; indeed the presence of the on-site Coulomb repulsion
$U$ has two main effects: i) it can change the eigenvalues and
their degeneracy; ii) it can lead, according to how the SS are
actually realized in terms of physical species, to quite
different shapes of the ground state phase diagram as a function
of $U$, even for models
of the same type (see for instance \cite{DOMO3} for the $BF$-type).\\

We just wish now to provide here a concrete example of the use of
the scheme proposed in this section, by presenting results on the
ground state of the models (\ref{H_i_prime}) with
${\mathcal{H}}^{(i)}= {\mathcal{H}}^{(b)}$ (subgroup 2), and
$s_1=-s_2=-1$; the coupling constants of these two models
explicitly read $X=t \, ; \, \tilde{X}=(1-\sigma) t \, ; \,
Y=\sigma t \, ; \, W=V=0 \, ; Q=-2 P =2t$; $U$ and $\mu$ can be
taken as arbitrary, as observed above. The part
${\mathcal{H}}^{(i)}$ is a GP for 3 Sutherland's species, namely
$F=\{ |\uparrow \rangle ,|\downarrow \rangle \}$,
$B_1=|0\rangle$, and $B_2=|\downarrow \uparrow\rangle$. Using the
extension of Sutherland's theorem one can derive the structure of
the ground state phase diagram as a function of $U$ and $n$;
details of this derivation can be found in\cite{DOMO3}. The
result is presented in fig.\ref{sym_fig2}.
\\The structure is the same for both models $\sigma=\pm1$. Four
regions can be recognized: in region I the ground state is made
of doubly occupied and empty sites; the eigenvalue of the energy
(per site) reads: $\epsilon_0=(U/2-1)\, n$; the eigenvectors are
pure $\eta$-pair states; the behavior of the two-particle reduced
density matrix
\begin{equation}
(\rho_2)_{i,j} \, = \, \frac { {}_m \langle \Psi_{(\sigma)} |
c^{\dagger}_{i\uparrow} c^{\dagger}_{i \downarrow} c^{}_{j
\downarrow} c^{}_{j \uparrow} \, | \Psi_{(\sigma)} \rangle_m } {
{}_m\langle \Psi_{(\sigma)} | \Psi_{(\sigma)} \rangle_m }
\end{equation}
in this region is the following\cite{EKS,SCHAD}
\begin{equation}
(\rho_2)_{i,j} \, \stackrel{|i-j| \rightarrow
\infty}{\longrightarrow} \, e^{i \frac{\pi}{2} (1-\sigma)\,
(i-j)} \, n_{\uparrow \downarrow} (1-n_{\uparrow \downarrow})
\label{corr-pure}
\end{equation}
where $n_{\uparrow \downarrow}$ is density of doubly occupied
sites. As eq.(\ref{corr-pure}) shows, in region I one has a
Off-Diagonal Long-Range-Order in this state, which implies that
the model is superconducting in this region. It is also possible
to see that, with respect to the AAS model, where the pure
$\eta$-pair phase is degenerate in the momentum of pairs
\cite{SCHAD}, here the pair hopping term selects a specific
momentum. This momentum (which is in principle observable through
neutronic spectroscopy) is strictly related to the kind of
$so(4)$-symmetry allowed by the model; in particular, for the
case of $so(4)_{(+1)}$ ({\it i.e.} $\sigma=+1$ in
eqns.(\ref{group2})) the selected momentum is $q=0$, while for
the case of $so(4)_{(-1)}$ ({\it i.e.} $\sigma=-1$ in
eqns.(\ref{group2})) the pairs have momentum $q=\pi$. The EKS
model has $0$-momentum pairs (see \cite{EKS}).
\\In region II all the possible types of sites (empty, singly
and doubly occupied) in the ground state are present; the energy
reads
\begin{equation}
\epsilon_0=(1-\frac{U}{2}) \left(\frac{1}{\pi}\arccos(\frac{1}{2}-
\frac{U}{4})-n \right)-\frac{2}{\pi}\sqrt{1-\frac{1}{4}
(1-\frac{U}{2})^2} \quad ,
\end{equation}
and the eigenstate is mixed $\eta$-pair, constructed on the $|
\mbox{\it ref } \rangle=|U=\infty\rangle$ states. The correlation
function behaves in this case \cite{SCHAD} as:
\begin{equation}
(\rho_2)_{i,j} \, \stackrel{|i-j| \rightarrow
\infty}{\longrightarrow} \, e^{i \frac{\pi}{2} (1-\sigma)\,
(i-j)} \, \, n_{\uparrow \downarrow} \frac{(1-n_{\uparrow
\downarrow}-n_s)}{(1-n_s)^2} \,  \langle
(1-n_i)(1-n_j)\rangle_{U=\infty} \label{corr-mixed}
\end{equation}
where $n_s$ is the density of singly occupied sites. This region
is again superconducting and is particularly interesting in that
it survives up to positive values of $U$ (which are expected to
be more physically meaningful, because the electronic on-site
interaction should be repulsive, even if partially screened by
the phononic effective attraction). This feature is shared by the
other known models like the AAS and the EKS ones; however it must
be observed that, around half-filling, the region II of our
models raises up to $U^{max}=6 t$, the highest value of 1-D
exactly known models. Finally, region III-a is metallic; the
eigenvalue is $\epsilon_0=-2 \pi^{-1} \sin(\pi n)$, while the
eigenstates are those of the $|U=\infty\rangle$ model; region
III-b is the particle-hole transformed of region III-a
($\epsilon_0=+2 \pi^{-1} \sin(\pi n))$. At half-filling it is
possible to show that a charge gap $\Delta=U-6t$ exists, which
makes the model an Insulator.
\\This is the picture of the ground state; for excited states
one cannot make use of Sutherland's theorem and is reconducted to
the spectrum of a whole $B^2 F$ case. On the physical point of
view, when the temperature is turned on, thermal fluctuations are
expected to break the Long-Range-Order of the superconducting
regions, according to Mermin-Wagner theorem. Nevertheless a
Quasi-Long-Range Order can survive, because the decay of
$(\rho_2)_{i,j}$ can preserve a tail over
macroscopically observable  distances.\\\\
Finally, we want to briefly comment on the mutual relationship
between the 32 $so(4)$-invariant models ${\mathcal{H}}^{(i)}$.
Within each of the 4 subgroups, there are transformations mapping
some models into others: in particular a model
${\mathcal{H}}^{(i)}$ characterized by $(s_1,s_2,\sigma)$ in its
parameters -- see eqns.(\ref{group1})$\div$(\ref{group4})-- is
connected to the one with $(-s_1,-s_2,\sigma)$ through the
transformation $c_{i, s} \rightarrow (-)^{i} c_{i, s}$;
analogously, a model with $(s_1,s_2,\sigma)$ in its parameters
can be mapped into that with $(s_1,s_2,-\sigma)$ through $c_{i, s}
\rightarrow [1-(1-(-)^{i}) n_{i, -s}] c_{i, s}$. Also, the
transformation $c_{i, \downarrow} \rightarrow (-1)^i
c^{\dagger}_{i, \downarrow}$ maps the subgroup
${\mathcal{H}}^{(b)}$ into the ${\mathcal{H}}^{(c)}$.
\\An exhaustive classification of all these transformations is out of
the purposes of the present paper, and will be treated in a
forthcoming paper. Here we just want to emphasize the following
aspects. In the first instance the above transformations act {\it
differently} on neighboring sites; this implies that they are not
graded similarity transformations, because (in matrix
representation) they are of the form $A \otimes^s B$, and not $A
\otimes^s A$ (see section V). Indeed, by exploiting the matrix
representation developed in section II it is possible to see that
none of the above models is connected to any other through a
graded similarity transformation; therefore, as far as the
$\Rc$-matrix is concerned, they are all {\it independently}
integrable. Secondly, it is also possible to show that the models
with different values of the parameter $Q$ cannot be connected by
{\it any kind} of transformation (not only similarity). Thirdly,
it should be pointed out that, even if some transformation maps a
given model into another, it is not that obvious in general that
the {\it ground state} of the former is mapped into the ground
state of the latter. This confirms that, also for what concerns
their physical stability properties, all the above models deserve
a deep interest. \label{sec-9}
%%%%%%%%%%%%%%%%%%%%%%%%%%%%%%%%%%%%%%%
%%%%%%%%%%%%%%%%%%%%%%%%%%%%%%%%%%%%%%%
%%%%%%%%%%%%%%%%%%%%%%%%%%%%%%%%%%%%%%%
\section{Conclusions}
In the preliminary part of this paper we have clarified the
relationship between the two approaches to fermionic integrable
systems, by providing a systematic method to pass from fermionic
operators to matrix representation and from the ${\Bbb C}$-number
YBE to fermion~models. Further, we have proved general results
about the symmetries in integrable systems: we have treated the
symmetries of the YBE, showing that it is invariant under {\it
graded} similarity transformations. Also, we have proved, under
quite general hypothesis, that the symmetries imposed on the
$\Rc$-matrix directly reflect onto the whole set of the constants
of motion ${\mathcal{J}}_n$ that the QISM determines. The
complementary question whether the symmetries of the Hamiltonian
reflect on the $\Rc$-matrix has been widely developed with the
PRT. These two general results are shown to be very effective
when combined together; in particular we applied both of them to
the study of the extended Hubbard models (EHM). By making use of
the matrix representation discussed above, we found all the EHM
that fulfill different kinds of symmetries and supersymmetries
(`doubly occupied sites', $so(4)$, $gl(2,1)$, $so(5)$, $u(2,2)$),
also including in the generators possible relative signs between
two neighboring sites. In particular, for the 2-parameter
subclass with $gl(2,1)$~supersymmetry, we have provided a
geometrical construction expressing any $gl(2,1)$ model as a
linear combination of the EKS-model and a U-supersymmetric model.
Furtherly, by showing that most of the integrable known model
(such as the EKS and the U-supersymmetric and others) are
reproduced by first/second degree polynomial $\Rc$-matrices, we
deduce the symmetries of their constants of motion.
\\Finally, focusing on the case of $so(4)$-symmetries (whose interest
in condensed matter has been discussed), we exploited the PRT to
find all the EHM that are derivable from first degree polynomial
$\Rc$-matrices. The symmetries of the constants of motion of
these models have also been discussed. Further, by making use of
the Sutherland's species technique, we have proposed a general
scheme to derive the spectrum and other physical properties of
these 32 models; in particular it has been observed that, by
means of Sutherland's theorem, the ground state energy can be
determined for all of them. Finally, thanks to the $so(4)$
symmetry, the addition of arbitrary Coulomb repulsion and
chemical potential terms to these integrable models can be used
to implement the $\eta$-pairs construction. We proposed an
example of concrete application of the described techniques to
two of the 32 models, for the ground state of which we have
explicitly given  the eigenvalue, the eigenvector and the pair
correlation functions in the different region of the phase
diagram (fig.\ref{sym_fig2}).
\\We also point out that all the extended Hubbard models
investigated above can also be used as `bulk systems' to which one
can add appropriate boundary interaction terms; this would lead to
new results in the context of models with Kondo impurities (see
for instance \cite{HQZ}).
%%%%%%%%%%%%%%%%%%%%%
\section*{Acknowledgments}
In the memory of prof. A. Izergin from Steklov Institute in St.
Petersburg University (Russia), the authors would like to remind
him for his intellectual generosity and helpful and interesting
discussions held
during his staying in Florence (Italy).
%%%%%%%%%%%%%%%%%%%%%%%%%%%%%%%%%%%%%%%%%%%%%%%%%%%%%%%%

\newpage

\begin{figure}
%\epsfxsize 12cm
%\centerline{\epsfbox{sym_fig1.eps}}
\epsfig{file=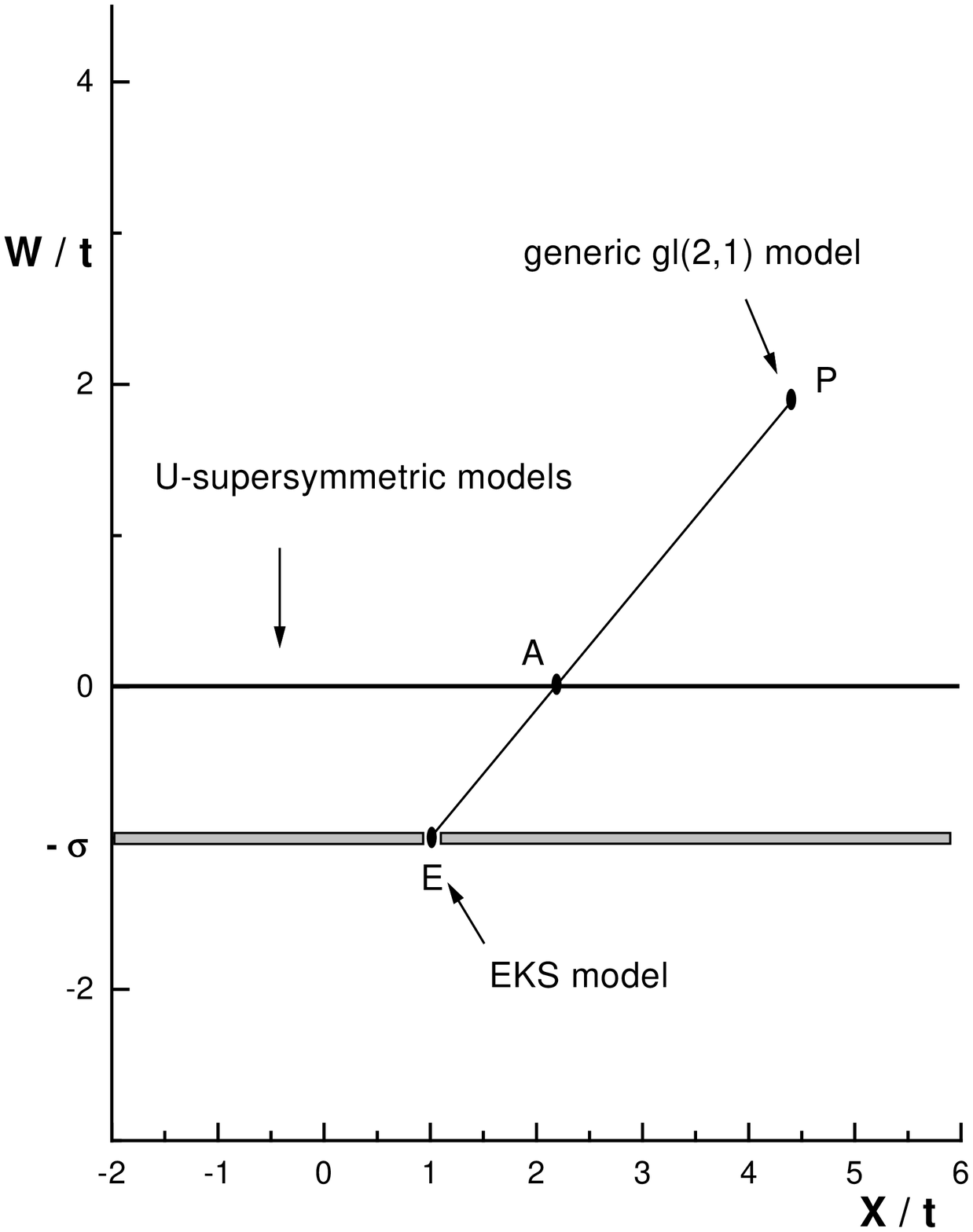,width=12cm}
\caption{The $X$-$W$ plane to describe the 2-parameter class of
$gl(2,1)$-invariant models. The coupling $X$ stands for correlated
hopping amplitude, while $W$ is the exchange term. The allowed
values are the two half planes $W/t >-\sigma $ and $W/t < -\sigma$,
with the addition of the single point $E=(1,-\sigma)$ that represents
the EKS models ($\sigma=\pm 1$).
The 1-parameter subclass of U-supersymmetric models are represented
by the $X$-axis ($W=0$).
\\Any $gl(2,1)$-invariant model (point $P$) can be written as a
linear combination of the EKS model and a particular U-supersymmetric
model, whose parameters can be geometrically determined by tracing
the line from $P$ to $E$, and finding the ascissa $X_A$ of the point
$A$ in which it intersects the $X$-axis.}
\label{sym_fig1}
\end{figure}
\begin{figure}
%\epsfxsize 12cm
%\centerline{\epsfbox{sym_fig2.eps}}
\epsfig{file=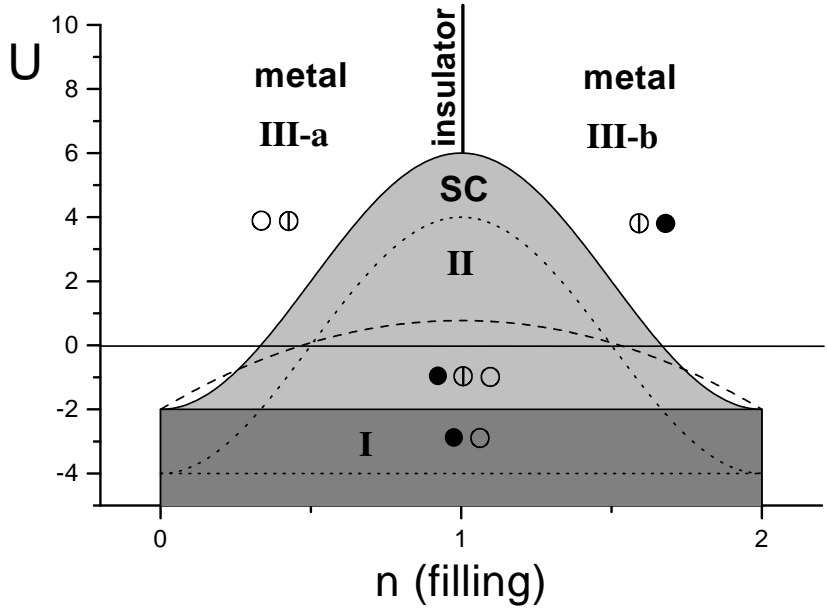,width=12cm}
\caption{Ground state phase diagram of the two models $X=1;
\tilde{X}=(1-\sigma); Y=-\sigma; P=-1; Q=2$ (subgroup 2,
eq.(\ref{group2})) from ref.[41].
The model exhibits an insulator-superconductor transition at $n=1$, for
$U_c=6$. The dashed line is
the EKS model, and the dotted line corresponds to the AAS model.}
\label{sym_fig2}
\end{figure}
\end{document}